\documentclass[10pt]{iopart}
\usepackage{iopams,setstack}  
\usepackage[ exponent-product = \cdot,
  inter-unit-product = \ensuremath{{}\cdot{}}
]{siunitx}
\usepackage{graphicx}
\usepackage{color}
\usepackage{cuted}
\begin{document}

\title[Plasmonic Purcell factor and coupling efficiency to surface plasmons.]{Plasmonic Purcell factor and coupling efficiency to surface plasmons. Implications for addressing and controlling optical nanosources}

\author{G. {Colas des Francs}, J. Barthes, A. Bouhelier, J.C. Weeber, A. Dereux}
\address{Laboratoire Interdisciplinaire Carnot de Bourgogne (ICB)
UMR 6303 CNRS-Universit\'e Bourgogne Franche-Comt\'e
9 Av. A. Savary, BP 47 870
F-21078 DIJON Cedex
FRANCE}
\ead{gerard.colas-des-francs@u-bourgogne.fr}
\author{A. Cuche, C. Girard}
\address{Centre d'Elaboration de Mat\'eriaux et d'Etudes Structurales (CEMES), CNRS, 29 rue J. Marvig,
Toulouse F-31055, France.}
\vspace{10pt}

\begin{abstract}
The Purcell factor $F_p$ is a key quantity in cavity quantum electrodynamics (cQED) that quantifies the coupling rate between a dipolar emitter and a cavity mode. Its simple form $F_p\propto Q/V$ unravels the possible strategies to enhance and control light-matter interaction. Practically, efficient light-matter interaction is achieved thanks to either i) high quality factor $Q$ at the basis of cQED or ii) low modal volume $V$ at the basis of nanophotonics and plasmonics. 
In the last decade, strong efforts have been done to derive a plasmonic Purcell factor in order to transpose cQED concepts to the nanocale, in a scale-law approach. In this work, we discuss  the plasmonic Purcell factor for both delocalized (SPP) and localized (LSP) surface-plasmon-polaritons and briefly summarize the expected applications for 
nanophotonics. On the basis of the SPP resonance shape (Lorentzian or Fano profile),  we derive closed form expression for the coupling rate to delocalized plasmons. The quality factor factor and modal confinement of both SPP and LSP are quantified, demonstrating their strongly subwavelength behaviour. 
\end{abstract}

%
%
\submitto{\JOPT}
%
%
\ioptwocol

\section{Introduction}
Nanophotonics permits light-matter interaction at the nanoscale, down to the single photon/single atom level. The motivations are notably sensitive sensing with applications such as nano-optical imaging (surface analysis), environnemental health (air pollutants, infectious agents detection), security (explosive detection) or healthcare (cancer early diagnosis, theranostics) and the miniaturization of photonics components for on chip integrated ultrafast devices.
The optical cross-section is a simple way to characterize the efficiency of light-matter interaction. For a molecule, it is typically $\sigma\simeq \SI{e-20}{\meter^2}$, that has to be compared to the focus area of a diffraction limited beam,  of the order of  $(\lambda/2)^2\simeq  \SI{e-13}{\meter \squared}$ in the visible domain. This unsuitability between the light confinement and the active size of the molecule points out the difficulty of so-called nanoscopy \cite{Hell:2007}. In the last decades, several strategies have been proposed to increase the efficiency of light-matter interaction. i) Increasing the absorption cross-section  by working at low temperature; indeed, in the limit of very low temperature ($T<\SI{10}{\kelvin}$), the molecular absorption cross-section increases up to the diffraction limit $\sigma_{0}=3\lambda^2/2\pi$ revealing that the molecule absorbs almost all the incoming light of a focused beam \cite{Celebrano-Sandoghdar:2011,Tamarat2000}. ii) Increasing the duration of the interaction by placing the molecule inside an optical microcavity presenting a high quality factor $Q$ \cite{Vahala:2003}. iii) Confining the excitation beam below the diffraction limit thanks to near-field optics \cite{Betzig1993,Veerman-Garcia-Kuipers-VanHulst:1999,Sick2001,OptExpMolendaGCF:2005,Wenger-Rigneault:2010} or plasmonics \cite{Brun-Drezet-Mariette-Chevalier-Woelh-Huant:2003,bharadwaj07a}.

Quantitatively, the efficiency of light-matter interaction can be inferred from the so-called cooperativity parameter $C$.  The meaning of this parameter is easily understood from a classical point of view \cite{Grynberg-Aspect-Fabre:2010,TanjiSuzuki-Vuletic:2011}. In free-space, the cooperativity can be expressed as the ratio between the resonant atomic cross-section $\sigma_0$ and the effective area $A_{eff}=\pi w_0^2$ of a gaussian beam with a beam waist $w_0$ ;  $C_0\approx \sigma_0/A_{eff}$. Therefore, the cooperativity quantifies the suitability between the focused spot and the molecule active area. In an optical Fabry-Perot cavity, it increases to $C=4C_0\cal F/\pi$ where $\cal F$ is the finesse of the cavity. The cavity enhances the free-space cooperativity by the number of wave round trips $\cal F/\pi$ inside the cavity and an additionnal factor of four accounting for the intensity enhancement at a mode antinode. In cavity quantum electrodynamics (cQED), the cooperativity writes for a single atom
\begin{eqnarray}
C=\frac{g^2}{2\kappa_{cav}n_1\Gamma_0}
\label{eq:Coop}
\end{eqnarray}
where $g$ is the coupling rate between the atom and the cavity mode. $\kappa_{cav}$ and $\Gamma_0$ refers to the cavity losses rate and the atom decay rate in vacuum, respectively ($n_1\Gamma_0$ is the decay rate in the homogeneous medium of optical index $n_1$). The strong coupling regime, $g\gg \kappa_{cav},\Gamma_0$ ($C\gg 1$) leads to a reversible energy exchange between the cavity and the atom. In the weak coupling regime, the cavity opens a new channel for the atom (irreversible) relaxation with a decay rate $\Gamma=(1+2C) \Gamma_0$. The Purcell factor quantifies the effect of the cavity on the atom decay rate and writes 
\begin{eqnarray}
F_p=\frac{\Gamma_{cav}}{n_1\Gamma_0}=2C
\end{eqnarray} 
with $\Gamma_{cav}=\Gamma-n_1\Gamma_0$ the modification of the decay due to the optical microcavity. It also expresses \cite{Purcell:1946}
\begin{eqnarray}
F_p=\frac{\Gamma_{cav}}{n_1\Gamma_0}=\frac{3}{4\pi^2}\left(\frac{\lambda_{em}}{n_1}\right)^3 \frac{Q}{V_{eff}}
\label{eq:Purcell}
\end{eqnarray} 
where $Q$ is the quality factor of the cavity and $V_{eff}$ the effective volume of the cavity mode involved in the coupling. $\lambda_{em}$ is the emission wavelength of the atom. This expression is equivalent to the classical description with the finesse for a Fabry-Perot cavity.

Remarkably, the Purcell factor expression (\ref{eq:Purcell}) points out that spontaneous emission can be efficiently controlled in an optical cavity presenting a high quality factor and/or a strongly confined mode. However, high Q cavities are obtained at the price of low modal (diffraction limited) confinement \cite{Vahala:2003}. In this context, molecular plasmonics proposes a new strategy for light--matter interaction \cite{Agio:2012,Sandoghdar-Agio:2012,Lodahl:2015}. The strong confinement of surface plasmon polaritons insures efficient coupling at a deeply subwavelength scale whereas cavity-QED increases the duration of interaction. At this point, we have to mention that the $Q$ factor entering the Purcell factor is the lower of the cavity factor and the atomic resonance. Actually, $1/Q=1/Q_{at}+1/Q_{cav}$ where  $Q_{at}$ and $Q_{cav}$ are the quality factor of the atom emission spectrum and cavity, respectively \cite{vanExter-Woerdman:1996}. That is why cQED generally works at low temperature where atomic resonance is sufficiently narrow so that cavity modifies the spontaneous emission ($Q_{at} \gg Q_{cav}$ and $Q\approx Q_{cav}$). On the contrary, low Q factor of plasmon resonance permits to work at room temperature and let envision high-speed optical devices \cite{Brinks-vanHulst:2013,Lalanne-Beveratos:2013}. This paves the way to ultrafast control at the nanoscale.

The Purcell factor describes the emitter-cavity coupling as a function of the optical cavity properties, independently of the emitter properties. Particularly, the best coupling efficiency is achieved for high $Q/V$ ratio, that occurs
either for a narrow resonance, or a deeply confined mode. The Purcell factor is therefore a key parameter to transpose cQED concepts to quantum plasmonics \cite{Agio:2012,Chang-Sorensen-Hemmer-Lukin:2006,Waks-Sridharan:2010,Buckley-Vukovic:2012,Hummer-Garciavidal:2013,Tame-Maier:2013,Hakami-Wang-Zubairy:2014,Torma-Barnes:2015,Rousseaux-GCF-Guerin:2015}. 

In this article, we first summarize the derivation of the Purcell factor in an optical microcavity with particular attention to the underlying hypothesis (section \ref{sect:Purcell}). In section \ref{sect:SPPcontrol}, we briefly discuss some expected applications of efficient emitter-SPP coupling. Then we study the concept of plasmonic Purcell factor. Since the mode confinement is the crucial parameter for achieving a high Purcell factor near a plasmonic nanostructures, we follow a progression from extended metal film (section \ref{sect:Mirror}) and plasmonic waveguide (section \ref{sect:SPPguide}) supporting delocalized surface plasmon polaritons (SPP) towards nanoparticles sustaining localized surface plasmons (LSP, section \ref{sect:LSP}).

\section{Purcell Factor}
\label{sect:Purcell}
\subsection{Purcell factor in an optical microcavity ; generalities}
The spontaneous emission at the angular frequency $\omega_{em}=2\pi c/\lambda_{em}$ from the excited state $\vert b \rangle$ to the ground state $\vert a \rangle$ of an excited atom presents a decay rate that follows Fermi's golden rule 
\begin{equation} 
\Gamma ({\bf r})=\frac{2\pi}{\hbar^2}\sum_{\mathbf{k}_n}\vert \langle a,\mathbf{k}_n\vert H_I \vert b,0 \rangle\vert^2 \delta(\omega_{em}-\omega_{\bf kn})
\label{eq:FermiRule}
\end{equation}
$H_I=-\hat{{\bf p}}\cdot \hat{{\bf E}}({\bf r})$ is the interaction hamitonian describing the coupling of an atom to an electromagnetic field within the dipolar approximation, taken at the position ${\bf r}$ of the atom.  The operators $\hat{{\bf p}}$ and $\hat{{\bf E}}$ refer to the atomic transition dipole moment and the electric field, respectively. In the following, we are interested in the effect of the cavity on the spontaneous emission rate. Therefore, we separate the decay rate into the free-space contribution (in an homogeneous medium of optical index $n_1$) and the cavity contribution
\begin{eqnarray}
n_1\Gamma_0&=&\frac{2\pi}{\hbar^2}\sum_{\mathbf{k}_n}\vert \langle a,\mathbf{k}_n\vert H_{0}\vert b,0 \rangle\vert^2 \delta(\omega_{em}-\omega_{\bf kn} )\\
&=&n_1\frac{p^2\omega_{em}^3}{3\pi \varepsilon_0\hbar c^3} \,, \text{and}
\label{eq:Gamma0} \\
\Gamma_{cav}&=&\frac{2\pi}{\hbar^2}\sum_{\mathbf{k}_n}\vert \langle g,\mathbf{k}_n\vert H_{cav}\vert e,0 \rangle\vert^2 \delta(\omega_{em}-\omega_{\bf kn}) \,.
\end{eqnarray}
In addition, we assume a single-mode cavity, resonant at $\omega_{c}$, therefore 
 \begin{eqnarray}
\Gamma_{cav}&=&\frac{2\pi}{\hbar^2}\vert \langle a,\mathbf{1}\vert H_{cav}\vert b,0 \rangle\vert^2 \delta(\omega_{em}-\omega_{c})
\label{eq:GammaDirac}
\end{eqnarray}
$\mathbf{1}$ indicates a photon into the cavity mode. The coupling rate $g$ between the atom and the cavity mode (see Eq.  \ref{eq:Coop}) obeys $\hbar g=  \vert\langle a,\mathbf{1}\vert H_{cav}\vert b,0 \rangle\vert$.
The electric-field operator associated to the single-mode cavity writes \cite{Grynberg-Aspect-Fabre:2010,Gerard:2003}
\begin{eqnarray}
\hat{{\bf E}}_{cav}({\bf r})=i\sqrt\frac{\hbar \omega_c}{2\varepsilon_0\varepsilon_1 V}{\bf f}({\bf r}) \hat{a} +h.c.
\label{eq:ElecQED}
\end{eqnarray}
where $V$ is the quantization volume,  $\hat{a}$ the boson operator and ${\bf f}({\bf r})$ describes the spatial variations of the mode into the cavity [$\vert{\bf f}({\bf r})\vert=0$ at a node and $\vert {\bf f}({\bf r})\vert=1$ at an antinode]. To achieve this expression, the classical electric field is expressed
\begin{eqnarray}
{\bf E}_{cav}({\bf r},t)=i\sqrt\frac{\hbar \omega_c}{2\varepsilon_0\varepsilon_1 V}{\bf f}({\bf r}) e^{-i\omega_c t}+c.c
\end{eqnarray}
and is normalized with respect to the energy 
\begin{eqnarray}
\hbar\omega_c&=&\frac{1}{2} \int [\varepsilon_0 \varepsilon({\bf r}) \mathbf{ { E}}^2(\mathbf{ r},t)  + \mu_0 \mathbf{ { H}}^2(\mathbf{ r},t) ] d{\bf r}\\
&=&\frac{\hbar \omega_c}{\epsilon_1 V} \int \varepsilon({\bf r}) \vert  {\bf f} (\mathbf{ r}) \vert^2  d{\bf r}\,,
\end{eqnarray}
where we assumed a non dispersive medium. Finally, the mode volume obeys  
\begin{eqnarray}
V=\frac{1}{\varepsilon_1}\int \varepsilon({\bf r}) \vert {\bf f} (\mathbf{ r}) \vert^2  d{\bf r} \,.
\label{eq:Vmode0}
\end{eqnarray}

\paragraph{Spectral shape of the resonance}
In the case of a lossy cavity, the dirac distribution in (\ref{eq:GammaDirac}) is replaced by the density of modes per unit angular frequency $N(\omega)$ 
(unit: $s.rad^{-1}$). 
Moreover, the profile of the mode resonance is assumed to be Lorentzian
\begin{equation}
\delta (\omega - \omega_c) \rightarrow N(\omega)=\frac{1}{\pi} \frac{\kappa_{cav}/2}{(\omega - \omega_c)^2 + \kappa_{cav}^2/4} \,.
\end{equation}
Defining the resonance quality factor $Q = \omega_c /\kappa_{cav}$ , we can rewrite 
\begin{equation}
N(\omega)=\frac{2 Q}{\pi \omega_c} ~~\frac{1}{1+ 4Q^2(\frac{\omega-\omega_c}{\omega_c})^2}.
\label{delta}
\end{equation}

\paragraph{Spatial profile}
Inserting the electric field operator (Eq. \ref{eq:ElecQED}) into the interaction hamiltonian, we achieve 
\begin{eqnarray}
\vert \langle g,\mathbf{1}\vert H_{cav}\vert e,0 \rangle\vert^2=\frac{p^2\hbar \omega_c}{2\varepsilon_0\varepsilon_1 V} \vert {\bf u}\cdot  {\bf f}(\mathbf{ r}) \vert^2
\label{eq:HintSpatial}
\end{eqnarray}
where we introduced the dipole moment orientation ${\bf u}$ ({\it i.e} ${\bf p}=p{\bf u}$).

\paragraph{Purcell factor}
Finally, using eq. (\ref{delta}) and (\ref{eq:HintSpatial}) the cavity contribution to the decay rate (eq. \ref{eq:GammaDirac}) simplifies to 

\begin{eqnarray}
\Gamma_{cav}&=&\frac{2p^2}{\hbar \varepsilon_0\varepsilon_1 } \vert  {\bf u}\cdot  {\bf f}(\mathbf{ r}) \vert^2 \frac{Q}{V} ~\frac{1}{1+ 4Q^2(\frac{\omega_{em}-\omega_c}{\omega_c})^2 }\,,
\end{eqnarray}
so that the normalized decay rate writes [using eq. (\ref{eq:Gamma0})  and $\lambda_{em} = 2\pi c/\omega_{em}$]
\begin{eqnarray}
\frac{\Gamma_{cav}}{n_1\Gamma_0}&=&\frac{3}{4 \pi^2}  \left ( \frac{\lambda_{em}}{n_1} \right )^{3}  \frac{Q}{V} ~\frac{\vert  {\bf u}\cdot  {\bf f}(\mathbf{ r})\vert^2}{1+ 4Q^2(\frac{\omega_{em}-\omega_c}{\omega_c})^2 } \,.
\end{eqnarray}

\subsubsection{Summarize}
The cavity contribution to the decay rate obeys 
\begin{eqnarray}
\label{eq:DecayDetuning}
\frac{\Gamma_{cav}}{n_1\Gamma_0}&=&F_p ~~\frac{\vert {\bf u}\cdot  {\bf f}(\mathbf{ r})\vert^2}{1+ 4~Q^2(\frac{\omega_{em}-\omega_c}{\omega_c})^2} \,,\text{with}\\
F_p &=&\frac{3}{4 \pi^2}  \left ( \frac{\lambda_{em}}{n_1} \right )^{3} ~ \frac{Q}{V} 
\nonumber
\end{eqnarray}
where $F_p$ is the so-called Purcell factor, ${\bf f}(\mathbf{ r})$ reveals the position dependency (from cancellation at a node to maximum effect at an antinode) and the denominator factor $[1+ 4Q^2(\omega_{em}/\omega_c-1)^2]$ shows the effect of the detuning between the emission frequency and the cavity resonance. 
Finally, the emitter couples to modes presenting a polarisation along the dipole moment [quantified by the term $\vert {\bf u}\cdot  {\bf f}(\mathbf{ r})\vert]$.

In the following, we are interested in defining the Purcell factor for a dipolar emitter coupled to a plasmonic nanostructure. It is therefore useful to recall the hypothesis done to demonstrate the Purcell factor expression (\ref{eq:Purcell}):

\begin{itemize}
\item the Purcell factor is associated to a given mode of the cavity  
\item the cavity resonance follows a Lorentzian shape
\item the mode volume can be estimated from expression (\ref{eq:Vmode0}), assuming a non dispersive medium. It equivalently writes 
\begin{eqnarray}
V&=&\frac{\int \varepsilon({\bf r}) \vert \mathbf{ { E}}(\mathbf{ r}) \vert^2  d{\bf r} }{Max\left[\varepsilon_1\vert \mathbf{ { E}}(\mathbf{ r}) \vert^2\right]}
\label{eq:Vmode}
\end{eqnarray}
where $\mathbf{  E}(\mathbf{ r})$ is the electric field associated to the cavity mode [${\bf f}(\mathbf{ r})=\mathbf{ { E}}(\mathbf{ r}) /Max\left[\vert \mathbf{ { E}}(\mathbf{ r})\vert \right]$ describes the mode profile]. 
\end{itemize}

\subsubsection{Purcell factor near a  nanofiber}
The full control of spontaneous emission in 3D optical cavity is a technological challenge and is bandwidth limited so that simpler configurations have been proposed. In particular, fluorescence emission into photonic nanowires can be enhanced over a large spectrum range so that it has been widely studied in the last decade \cite{Yalla-Hakuta:2012,Claudon-Gregersen-Lalanne-Gerard:13}.  To derive the Purcell factor near a waveguide, we define an arbitrary quantization length $L$. The electric-field operator and the density of mode write, respectively \cite{JunPhD:2010}
\begin{eqnarray}
\hat{{\bf E}}({\bf r})&=&i\sqrt\frac{\hbar \omega_c}{2\varepsilon_0\varepsilon_1 A_{eff}L}{\bf f}({\bf r}) \hat{a} +h.c. \;, \text{with}\\
A_{eff}&=&\frac{\int \varepsilon(x,z) \vert \mathbf{ { E}}(\mathbf{ r}_{xz}) \vert^2  dxdz }{Max\left[\varepsilon_1\vert \mathbf{ { E}}(x,z) \vert^2\right]}
\label{eq:Amode}  \;, \text{and} \\
 N(\omega)&=&\frac{L}{\pi} \frac{1}{v_g} \;, \text{with }  v_g=\frac{d\omega}{dk_g}
 \end{eqnarray}
$A_{eff}$ defines the mode effective area and $v_g$ is the group velocity of the guided mode.  $N^{2D}=1/\pi v_g$ is the density of guided modes. We then proceed as previously and the Purcell factor for the guided mode simplifies to  
\begin{eqnarray}
F_p=\frac{\Gamma_{guided}}{n_1\Gamma_0}=\frac{3}{4 \pi}  \frac{(\lambda_{em}/n_1)^2}{A_{eff}} \frac{n_g}{n_1} \;.
\label{eq:Purcell2D}
\end{eqnarray}
The Purcell factor near a photonic nanofiber is governed by the group index $n_g=c/v_g$ of the guided mode and its transverse confinement $A_{eff}$. High Purcell factor necessitates low group velocity (\textit{e.g.} near the band-edge of the dispersion relation \cite{Woldeyohannes-John:2003,Chen-Brandes:2009})  and/or a highly confined mode.

\subsubsection{Purcell factor in a Fabry-P\'erot cavity}
We finally consider the one-dimensional (1D) planar waveguide. The density of guided modes obeys $N^{1D}=\omega/2\pi n_{eff} n_g$ where $n_{eff}$ refers to the mode effective index. The Purcell factor becomes
\begin{eqnarray}
\label{eq:Purcell1D}
F_p&=&\frac{\Gamma_{guided}}{n_1\Gamma_0}=\frac{3}{4}  \frac{(\lambda_{em}/n_1)}{L_{eff}} \frac{n_{eff} n_g}{n_1^2} \;, \text{with}\\
L_{eff}&=&\frac{\int \varepsilon(z) \vert \mathbf{ { E}}(z) \vert^2  dz }{Max\left[\varepsilon_1\vert \mathbf{ { E}}(z) \vert^2\right]}
\label{eq:Lmode}  
\end{eqnarray}
$L_{eff}$ is the mode effective length and characterizes its confinement. 

\subsubsection{Coupling efficiency ($\beta$-factor)}
So far, we have introduced the Purcell factor that quantifies the coupling strength between a dipolar emitter and a photonic structure. Since the emitter could relax to its ground state thanks to various channels  (cavity mode, leakage, non radiative energy transfer), it is also useful to define the coupling efficiency to the cavity mode. It is the so-called $\beta$-factor
\begin{equation}
\beta = \frac{\Gamma_{cav}}{\Gamma_{tot}} = \frac{\Gamma_{cav}}{\Gamma_{cav} + \Gamma_{other}} 
\end{equation}
where $\Gamma_{tot}=\Gamma_{cav} + \Gamma_{other}$ is the total decay rate that includes all the relaxation channels. In a single mode cavity, the relaxation channels are generally the cavity mode with the decay rate $\Gamma_{cav}$, and coupling to leaky modes. 
Usually, leakage are of the same order of magnitude than the initial radiative rate ($\Gamma_{other} \simeq n_1 \Gamma_0$) so that  $\beta~\simeq~\Gamma_{cav}/(\Gamma_{cav}+n_1\Gamma_{0})~\simeq~F_p / (1 + F_p)$. For instance, a Purcell factor of about $F_p\sim 10$ corresponds to a coupling efficiency $\beta \sim 90 \%$. High $\beta$ factor is required for low threshold lasering \cite{Nomura-Arakawa:2008}.

\subsection{Decay rate near lossy and dispersive materials}
In presence of a lossy and dispersive medium, the modification of the decay rate can be described either within classical Lorentz
model of an oscillating dipole \cite{metiu84,Girard1995l,JCPGCF:2002} or within the full quantum description \cite{Knoll-Scheel-Welsch:01}. In both cases, this leads to the following expression for the rate modification 
\begin{equation} 
\frac{\Gamma_u (\bf r)}{n_1\Gamma_0}=\frac{6\pi}{n_1k_0}Im [G_{uu}(\mathbf{r},\mathbf{r},\omega_{em}) ]
\label{eq:DecayGreen}
\end{equation}
where $\mathbf{G}$ is the Green's tensor associated to the emitter surroundings and $G_{uu}$ refers to its diagonal components along the dipolar direction $\mathbf u$. This expression quantifies the modification of the decay rate in a complex surroundings and will be our starting point to determine the effect of plasmonic nanostructures on the fluorescence decay rate. 
The expression (\ref{eq:DecayGreen}) is a generalization of the Fermi's golden rule to dispersive and lossy surroundings. Indeed, the Fermi's golden rule (\ref{eq:FermiRule}) can be recast in the form 
\begin{eqnarray}
\Gamma ({\bf r})&=&2\pi g^2({\bf r},\omega_{em}) N(\omega_{em}) \;,\\
\hbar g({\bf r},\omega_{em})&=& \vert\langle a,\mathbf{1}\vert H_{cav}\vert b,0 \rangle\vert \\
&=& \sqrt{\frac{\hbar \omega_c}{2\varepsilon_0\varepsilon_1 V}} \vert {\bf p} \cdot  {\bf f}(\mathbf{ r}) \vert
\end{eqnarray}
where we introduced the coupling strength $g$ and density of modes $N(\omega)$, as discussed above. It is well-known from scattering formalism that the partial local density of states (P-LDOS) is related to the Green's dyad (unit: $s.m^{-3}$) \cite{JCPGCF:2002,Vries1998} 
\begin{eqnarray}
\rho_u({\bf r},\omega_{em})=\frac{k_0^2}{\pi\omega }Im [G_{uu}(\mathbf{r},\mathbf{r},\omega_{em}) ] 
\end{eqnarray}
so that Eq. (\ref{eq:DecayGreen}) and (\ref{eq:FermiRule}) are fully equivalent in a non-absorbing medium. The P-LDOS also writes from cQED considerations 
\begin{eqnarray}
\rho_u({\bf r},\omega)=\frac{f_u({\bf r})}{2\epsilon_1V}N(\omega) .
\end{eqnarray}
Therefore, the P-LDOS grasps the spatial dependency of the coupling to the cavity mode by $f_u({\bf r})$ and the density of available modes for light emission $\rho(\omega)=N(\omega)/V$,  per unit angular frequency and unit volume.

\section{Plasmonic addressing and control of optical nanosources}
\label{sect:SPPcontrol}
Before discussing the plasmonic Purcell factor, it is worthwile to briefly introduce some expected applications of coupled emitter-SPP configurations. We indicate here the main applications and refer the reader to the literature for more complete descriptions.  
\subsection{Surface enhanced spectroscopies}
Surface Enhanced Raman Spectroscopy (SERS) is probably among the first application of molecular material coupled to plasmonics nanostructures \cite{moskovits05} and is now available at the single molecule level \cite{LeRu-Etchegoin:12} so that ultrasensitive chemical or biosensors  are expected \cite{LeRu-Etchegoin:08}. In the context of Purcell factor, we would like to mention that the SERS efficiency follows a $\propto Q^2/V$ law with the LSP properties \cite{Maier:2006}.  

Following SERS, metal enhanced fluorescence (MEF) leads to enhancement factor of the order of tens \cite{Fort-Gresillon:2005,Lakowicz2005,Kuhn-Sandoghdar:2006,Anger-Bharadwaj-Novotny:2006,Viste-Plain:10,Derom-GCF:2013} with possible applications to nanotheranostics \cite{loo04,Wang-Chuang-AnnieHo:12}. The fluorescence increase results from excitation field enhancement and emission rate modification (Purcell effect). However, due to the non radiative energy transfer to metal nanostructures, it is crucial to distinguish the radiative from non radiative rates. At the end, the critical  parameter is the quantum yield of the emitter so that the total enhancement is generally limited by the intrinsic quantum yield of the fluorescent molecule \cite{Brokman-Hermier:2004,Buchler-Kalkbrenner-Hettich-Sandoghdar:2005,Mertens-Koenderink-Polman:2007,Carminati-deWilde:2015}.  Remarkably, coupling an emitter to a plasmonic nanostructures opens the way to the control of the photophysical processes \cite{CPLGirardGCF:2005}, notably blinking effect \cite{stefani07,fu07,Greffet-Dubertret:2015} and photobleaching \cite{Cang-Xhang:2013}, but also to modify the ratio between magnetic and electric allowed dipolar transitions \cite{Karaveli-Zia:2011,Taminiau-Zia:2012}.

\subsection{Nano-optical antennas}
Since a plasmonics nanostructure efficiently interfaces a single molecule to far-field radiation \cite{HechtScience:2005,Taminiau-Segerink-vanHulst:2007,PRBHuang:2008,Bonod2010} the concept of optical nano-antenna has emerged since a decade \cite{Bharadwaj-Deutsch-Novotny:2009,Greffet-Laroche-Marquier:2010,Krasnok-Belov:2015}. Optical nano-antennas rely on plasmonics nanostructures to efficiently redirect the fluorescence emission and cQED-like description gives an insight of the coupling mechanism \cite{Agio:2012,Delga-GarciaVidal:2014}. Recently, a Purcell factor up to 1000  keeping a reasonable quantum yield and with a collection efficiency of 84\% was demonstrated \cite{Akselrord-Mikkelsen:2014}. Moreover, optical nano-antenna could efficiently interface molecular fluorescent emission and a nanophotonic waveguide \cite{Koenderink:12} with possible applications to realize a platform for quantum optics. In addition, coupling a single photon source to an optical nano-antenna permits to control its emission cadency  \cite{Schietinger-Barth-Aichele-Benson:2009,Marty-Arbouet-Paillard-Girard-GCF:2010,MallekZouari:2010,Celebrenao-Sandoghdar:2010,Cuche2010,Busson-Bonod-Bidault:2012,Mollet2012,GU-MARTIN:2012,Rockstuhl:2014,Nerkararyan-Bozhevolnyi:2014}. Realization of indistiguishable single photons is also a major issue \cite{Beveratos:2014,Grange-Auffeves:2015}.

\subsection{SPP amplification and SPASER}
Taking advantage of the analogy between optical microcavities and plasmonic nanostructures, the concepts of plasmon nanolaser and amplifier were proposed \cite{Bergman2003,Protsenko:2005,Berini-deLeon:2012}. It consists of a gain medium in contact to a metal nanostructure so that stimulated emission of plasmon occurs. The efficiency of the stimulated emission strongly depends on the Purcell factor \cite{Winter-Wedge-Barnes:2006,deLeon-Berini:2008,ColasdesFrancsOptExp:2010,GrandidierJMicrosc:2010}.

\subsection{Dipole-dipole coupling}
Finally, since an emitter can be efficiently coupled to a surface plasmon, it has been proposed to use plasmon to couple two emitters for applications such as long range resonant energy transfer (above 10 nm) \cite{Andrew-Barnes:2000,MartinCano-MartinMoreno-GarciaVidal-Moreno:2010,ZuritaSanchez:2013,Karanikolas-Bradley:2014,Krachmalnicoff:2015} or qubits entanglement \cite{Fleischhauer:2010,GonzalezTuleda-GarciaVidal:2011}. 

\bigskip
In the following, we discuss the plasmonic Purcell factor for typical configurations. We first consider the well-known case of extended metal film that is the simplest case of delocalized SPP enabling to clearly identify the coupling mechanisms \cite{ford84,Barnes1998b}. 

\section{Quantum emitter decay rate near a metal mirror}
\label{sect:Mirror}
\subsection{Thick mirror}
\begin{figure}
\includegraphics[width=8cm]{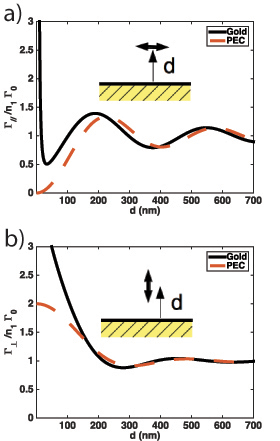}
\caption{Decay rate as a function of the distance to a gold or PEC surface. a) Dipole moment parallel to the surface. b) Dipole moment perpendicular to the surface. The emission wavelength is  $\lambda_{em}=\SI{670}{\nano \meter}$.{The gold permittivity is taken from tabulated data \cite{Johnson-Christy:1972}.}}
\label{fig:GamaSlab}
\end{figure}

Figure \ref{fig:GamaSlab} presents the dipolar total decay rate as a function of the distance to a gold surface \cite{Chance:1975}. The perfect mirror (perfect electric conductor - PEC) case is also shown for comparison \cite{Hinds:1991}. Far from the surface, we observe a typical interference pattern since the driving reflected field has to be in phase with the dipolar oscillation to enhance the emission rate. When the emitter touches the surface, the decay rate presents a finite value for the perfect mirror. It fully cancels for a dipole parallel to the surface (Fig. \ref{fig:GamaSlab}a) whereas it doubles for a perpendicular dipole (Fig. \ref{fig:GamaSlab}b), in agreement with the image dipole induced into the conductor. In case of real metal, the behavior is rather different close to the surface due to the apparition of new decay channels such as excitation of SPP and non radiative energy transfer to the metal film\cite{ford84,Barnes1998b}.

\begin{figure}
\includegraphics[width=8cm]{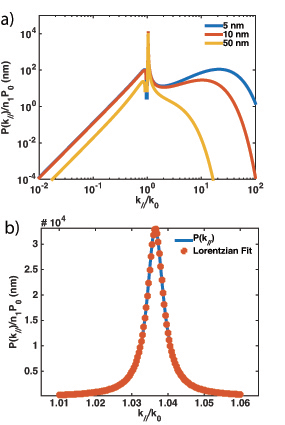}
\caption{a) Dipolar emission as a function of the wavector parallel to the surface $k_{\parallel}$. b) Resonant behaviour of the Au/air SPP. The Lorentzian fit is peaked at the SPP effective index $n_{SPP}=1.036$ and with a FWHM $n_{SPP}^{''}=\SI{2.70e-3}{}$ ($L_{SPP}=\lambda_{em}/4\pi n_{SPP}^{''}=\SI{20}{\micro\meter}$). The dipole is perpendicular to the surface. The emission wavelength is  $\lambda_{em}=\SI{670}{\nano \meter}$. Direct (free-space) dipolar emission is not included in $P(k_{\parallel})$. }
\label{fig:PSlab}
\end{figure}

\subsubsection{Relaxation channels}
The various contributions to the total decay rate are easily determined from the Sommerfeld expansion of the dipolar emission, represented in Fig. \ref{fig:PSlab} \cite{Barnes1998b,Chance:1975}. Indeed, since the metal/air interface is invariant along $\mathbf {r_{\parallel}}=(x,y)$, it is possible to expand the 
Green's dyad over the wavenumber $k_{\parallel}$ so that the  total decay rate (Eq. \ref{eq:DecayGreen}) obeys (see also \ref{sect:G1D})
\begin{eqnarray}
\frac{\Gamma_u(d)}{n_1\Gamma_0}&=&\int_0^\infty \mathcal{P} (k_{\parallel})dk_{\parallel} \;,\text{with} \\
\mathcal{P} (k_{\parallel})&=& \frac{P(k_{\parallel})}{n_1P_0}
\label{eq:GammaSpec}
\end{eqnarray}
where $u=\parallel, \perp$ represents the orientation of the emitter and $d$ is the distance to the metal surface. $P_0=\omega_{em}^4\vert p\vert^2/12\pi \epsilon_0 c^3$ is the power radiated by the oscillating dipole in vacuum. $P(k_{\parallel})$ is the dipolar emission power spectrum (in the $k_{\parallel}$-space) and represents the dipolar emission at a surface wavector $k_{\parallel}$. $\mathcal{P} (k_{\parallel})$  is $P(k_{\parallel})$, normalized with respect to $P_0$ (unit of $\mathcal{P}$: meter). 

It is useful to distinguish the radiative waves for which $k_{\parallel}\le n_1k_0$ that contributes to the radiative rate $\Gamma_{rad}$ and the evanescent waves $k_{\parallel}> n_1k_0$. Evanescent waves corresponds to either SPP or lossy waves (LSW) \cite{ford84,Barnes1998b}. Particular attention has to be paid to SPP contribution, shown in Fig. \ref{fig:PSlab}b. It presents a Lorentzian profile that permits to derive a closed form expression for the Purcell factor as we will discuss in details later. Finally, the large wavectors are associated to electron scattering losses. These so called lossy waves are responsible for fluorescence inhibition close to the metal surface. Let us mention that an additional non radiative energy transfer, namely electron-hole pair creation in the metal, could also occur at very short separation distances ($d<\SI{1}{\nano\meter}$) but is not included in this model since it involves non local description of the metal dielectric constant\cite{ford84}.

\begin{figure}
\includegraphics[width=8cm]{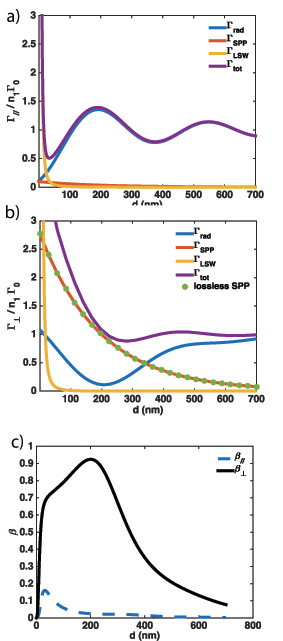}
\caption{Contributions to the total decay rate as a function of the distance to the gold surface for a dipole parallel (a) or perpendicular (b) to the surface. c) Coupling efficiency to the SPP. The emission wavelength is  $\lambda_{em}=\SI{670}{\nano \meter}$. {In b), the green dots represents the SPP contribution calculated assuming a lossless metal.}}
\label{fig:GamaContrib}
\end{figure}

Eventually, we present in Fig. \ref{fig:GamaContrib} the radiative, SPP and lossy waves contribution to the total decay rate. The radiative decay rate is the main channel for large separation distances whereas lossy waves dominate very close to the metal surface. The SPP contribution is practically negligible for a parallel dipole (Fig. \ref{fig:GamaContrib}a) since Au/air SPP is TM polarized (that is the electric field is perpendicular to the metal surface), hence low $\beta_{\parallel}$ factor (Fig. \ref{fig:GamaContrib}c). At the opposite, we observe that a perpendicular dipole is efficiently coupled to a SPP (Fig. \ref{fig:GamaContrib}b).  About 90\% of the dipolar emission couples to the Au/air SPP for a separation distance of  $d=\SI{200}{\nano\meter}$, see Fig. \ref{fig:GamaContrib}c. This high $\beta$-factor originates from a low radiative rate due to a destructive interference between the direct and reflected dipolar fields (Fig. \ref{fig:GamaContrib}b). At shorter distances where the SPP rate is higher,  we achieve $\beta_{\perp}=70\%$ for $d=\SI{40}{\nano\meter}$, see Fig. \ref{fig:GamaContrib}c.

At this point, a closed form expression of the plasmonic Purcell factor is achievable. Since the emitted power follows a Lorentzian profile near the SPP resonance, the integration over the SPP contribution leads to  \cite{Barthes-GCF-Bouhelier-Weeber-Dereux:2011}
\begin{eqnarray}
\label{eq:SPPPurcell1}
\frac{\Gamma_{SPP}}{n_1\Gamma_0}&=&\frac{\pi}{2}\frac{\mathcal{P}(k_{SPP})}{L_{SPP}}
\end{eqnarray}
This extends the Purcell factor definition to a (lossy) SPP. Care has to be taken when interpretating plasmonic Purcell factor, in particular the role of losses. {We derive in \ref{app:Lossless} the SPP contribution assuming a lossless metal. It perfectly matches the SPP contribution calculated for a real lossy metal (see green dots in Fig. \ref{fig:GamaContrib}b).} So the plasmonic Purcell factor does not depend on the propagation length \cite{Barthes-GCF-Bouhelier-Weeber-Dereux:2011,Jung:2012}, although it explicitly appears in the denominator of expression  (\ref{eq:SPPPurcell1}). Mathematically, the integral of the Lorentzian resonance gives the number of  supported modes and does not depend on ohmic losses \cite{ColasdesFrancsPRB:2009}. This is finally not surprising since SPP rate defines the coupling efficiency to the propagating SPP, no matter of how energy is dissipated afterward.   

\begin{figure}
\includegraphics[width=8cm]{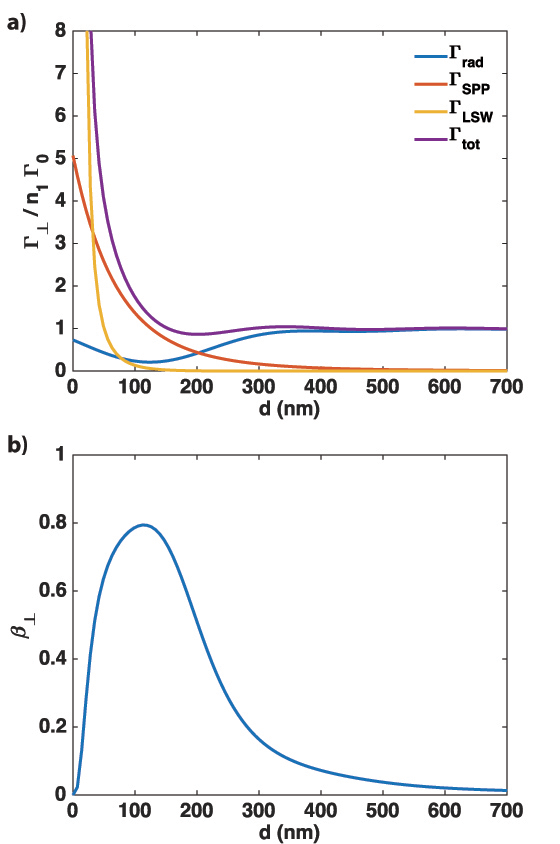}
\caption{{a) Contributions to the total decay rate as a function of the distance to the gold surface. b) Coupling efficiency to the SPP. The emission wavelength is  $\lambda_{em}=\SI{525}{\nano \meter}$ and the dipolar emitter is perpendicular to the surface.}}
\label{fig:GamaContrib525}
\end{figure}

As a consequence, the Purcell factor as well as the coupling efficiency to a SPP can be high, even in presence of strong losses. We plot in Fig. \ref{fig:GamaContrib525} the different contributions to the decay rate at $\lambda_{em}=\SI{525}{\nano \meter}$.  {Due to strong losses in gold, it is difficult to separate the SPP and LSW contributions to the total decay rate. This behaviour is discussed in detail in \ref{app:lossy}. For simplicity, we estimate the SPP decay rate from the emitter power integrated over $1<k_{\parallel}/k_0<1.4$}. All decay channels present a behaviour very similar to the $\lambda_{em}=\SI{670}{\nano \meter}$ case (compare with Fig. \ref{fig:GamaContrib}). Although we expect strong losses in the metal due to interband transitions, we still observe $\beta_{\perp}=80\%$ for $d=\SI{100}{\nano\meter}$, see Fig. \ref{fig:GamaContrib525}b. The SPP has an effective index $n_{SPP}=1.10$ (effective wavelength $\lambda_{SPP}=\SI{477}{\nano\meter}$) but an extremelly short propagation length $L_{SPP}=\SI{640}{\nano\meter}$. Therefore, the SPP presents only a single spatial oscillation over its propagation length. This quasi-mode, although efficiently excited would not be of interest for the control of a dipolar emission. 

\begin{figure}
\includegraphics[width=8cm]{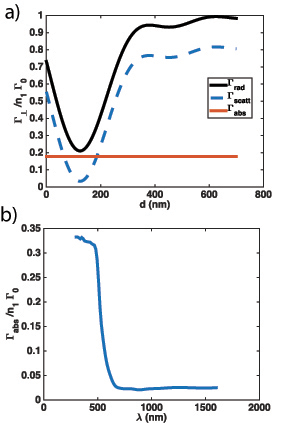}
\caption{a) Radiative and scattering rates at the emission wavelength $\lambda_{em}=\SI{525}{\nano \meter}$. b) Absorption rate ($\Gamma_{abs}=\Gamma_{rad}-\Gamma_{scatt}$) as a function of the wavelength. The dipole is perpendicular to the surface.}
\label{fig:GamaAbs525}
\end{figure}
So far, we only discussed the emission process of the dipolar emitter. Another quantity of interest is the collection efficiency. We plot in figure \ref{fig:GamaAbs525} the scattering rate $\Gamma_{scatt}$, defined as the dipolar power dissipated in the far-field zone. The difference with the radiative rate $\Gamma_{rad}$ is due to absorption in the metal but doesn't depend on the distance to the metal surface \cite{Chance:1975} {(see also \ref{app:Gscatt})}. We observe in Fig. \ref{fig:GamaAbs525}b that absorption becomes negligible above $\lambda>\SI{650}{\nano\meter}$.

\begin{figure}
\includegraphics[width=8cm]{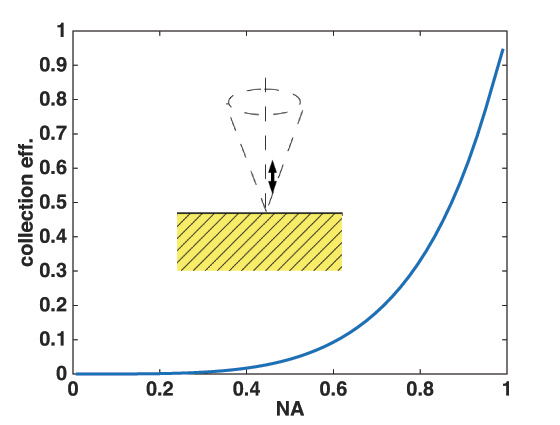}
\caption{Collection efficiency as a function of the detection numerical aperture (NA). The emission wavelength is  $\lambda_{em}=\SI{670}{\nano \meter}$ and the dipole is perpendicular to the surface.}
\label{fig:eta670}
\end{figure}
As last quantity, we plot in Fig. \ref{fig:eta670} the collection efficiency $\eta=\Gamma_{scatt}(NA)/\Gamma_{rad}$ as a function of the detection numerical aperture (NA). $\Gamma_{scatt}(NA)$ refers to the power scattered in a given NA. A NA=0.6 air objectif collects 10\% of the emitted signal so that strategies has to be developed to improve this efficiency such as surface plasmon coupled emission (SPCE) \cite{Lakowicz2005} or grating decoupler \cite{Aouani-Wenger:2011,Choy-Loncar:2013,Kumar-Dubertret-GCF:2015}.  

\subsubsection{Quality factor and mode confinement}
\label{sect:Qspp-Leff}
Micro-optical cavities are generally characterized by their mode confinement and quality factor. For comparison purposes, it is therefore convenient to estimate also SPP confinement and quality factor. 

Firstly, the SPP quality factor is estimated from the resonance Lorentzian profile (Fig. \ref{fig:GamaContrib}b). At $\lambda_{em}=\SI{670}{\nano \meter}$ it comes 
$Q=k_{spp}/\Delta k_{spp}=n_{spp}/2n''_{spp}=192$. 

Secondly,  we would like to characterize the mode confinement. For this purpose, we identify the  SPP rate $\Gamma_{SPP}/n_1\Gamma_0$ (Eq. \ref{eq:SPPPurcell1}) to the Purcell factor (Eq. \ref{eq:Purcell1D}). Remembering the hypothesis done to achieve the Purcell factor expression, we calculate  $\Gamma_{SPP}$ at the gold/air interface where the SPP field amplitude is the highest  and sum the contributions for dipoles along the three directions (so that $\vert \bf u \cdot {\bf f}\vert =1$). Indeed, the decay rate obeys then
\begin{eqnarray}
&&\frac{\Gamma_{cav,x}}{n_1\Gamma_0}+\frac{\Gamma_{cav,y}}{n_1\Gamma_0}+\frac{\Gamma_{cav,z}}{n_1\Gamma_0}=F_p \,,\\
&&\equiv\frac{3}{4}  \frac{(\lambda_{em}/n_1)}{L_{eff}} \frac{n_{eff} n_g}{n_1^2} \,.
\label{eq:Purcell1DIdent}
\end{eqnarray} 
We estimate the SPP effective and group indices $n_{spp}=1.036$ and $n_g=1.16$, respectively,  from the dispersion relation. From identification to the Purcell factor (Eq. \ref{eq:Purcell1DIdent}), we can attribute the effective length $L_{eff}=\SI{203}{\nano \meter}\simeq 0.6(\lambda_{spp}/2)$ to the gold/air SPP (Purcell factor of $F_p=2.9$).  This effective length is  similar to the SPP penetration depth in air $\delta/2=\lambda/2k_0\sqrt{n_{spp}^2-1}=\SI{196}{\nano \meter}$. Indeed, neglecting the mode extension in gold, its effective length can be estimated as (assuming the validity of this expression in presence of the absorbing and dispersive gold mirror)
\begin{eqnarray}
L_{eff}&=&\frac{\int \vert \mathbf{ { E}}(z) \vert^2  dz }{Max\left[\vert \mathbf{ { E}}(z) \vert^2\right]} 
=\int_0^\infty e^{-2z/\delta} dz= \frac{\delta}{2}
\end{eqnarray}
and $\delta$ is a good parameter to estimate the SPP confinement.  

For $\lambda_{em}=\SI{525}{\nano \meter}$, we achieve $Q=8$ and 
$L_{eff}=\SI{223}{\nano \meter}\simeq (\lambda_{spp}/2)$ (and $\delta/2=\SI{235}{\nano \meter}$). However, plasmonic Purcell factor is menaningless in this spectral range as pointed out in the previous section.

\subsection{Thin metal film}
\subsubsection{Relaxation channels}
\begin{figure}
\includegraphics[width=8cm]{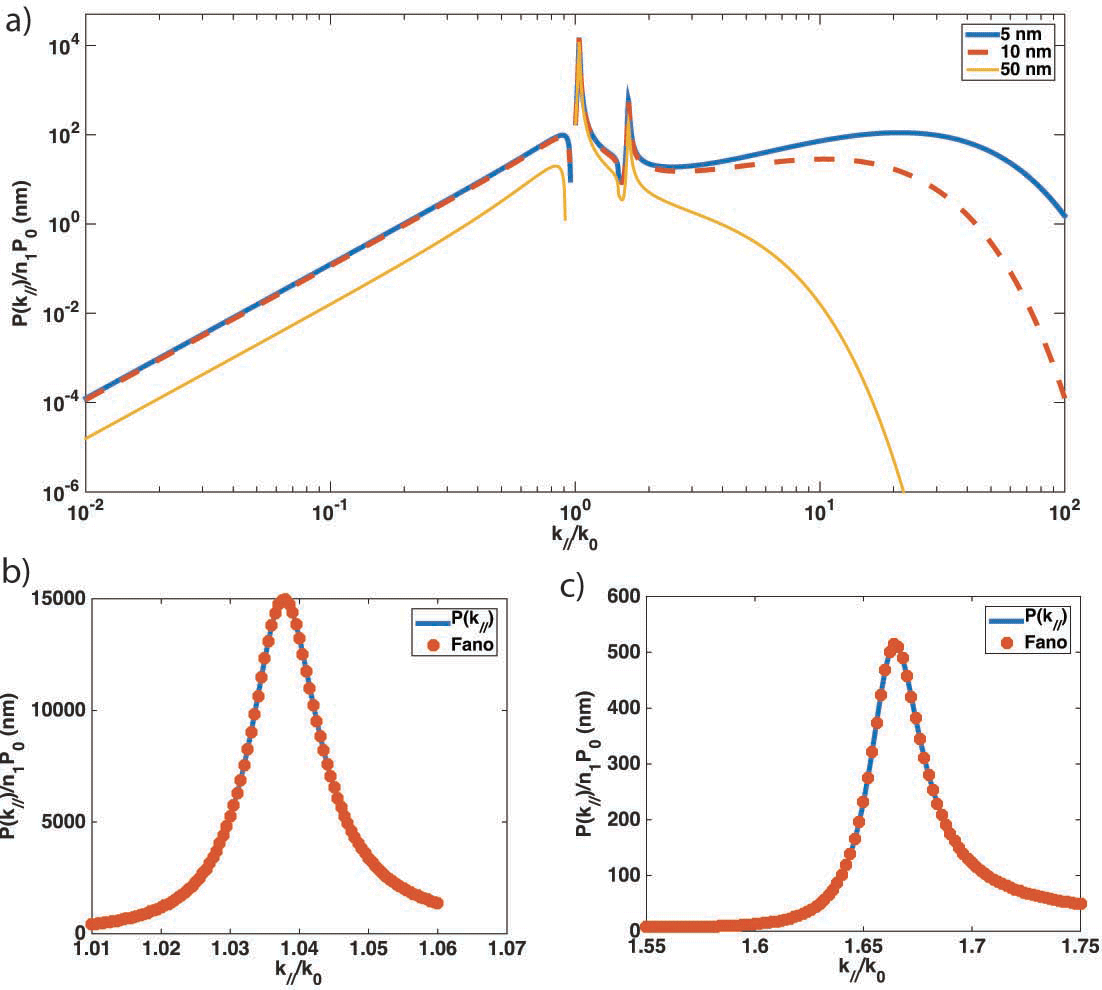}
\caption{a) Dipolar emission power spectrum $P(k_{\parallel})$ above a 50 nm thin gold film deposited on a glass substrate (optical index $n_{sub}=1.51$). b) Resonant behaviour of the Au/air SPP (d=10 nm, blue solid line). The Fano parameters (red dots) are $n_{spp1}=1.0375$, $n_{SPP1}^{''}=\SI{6.07e-3}{}$, $q_1=50$ and $b_1=-0.05$. c) Resonant behaviour of the Au/glass SPP (d=10 nm). The Fano parameters are $n_{spp2}=1.6627$, $n_{SPP2}^{''}=0.0145$, $q_2=4.12$ and $b_2=16.4$. The dipole is perpendicular to the surface. The emission wavelength is  
$\lambda_{em}=\SI{670}{\nano \meter}$.}
\label{fig:GamaThinSlab}
\end{figure}
We now turn to the thin metal film case. We consider a 50 nm gold film deposited on a glass substrate, and an emission wavelength $\lambda_{em}=\SI{670}{\nano \meter}$. The Au/air and Au/glass SPP modes couple and to form a leaky and a bound SPP.  We determine their characteristics using the reflection pole method \cite{Anemogiannis-Glytsis-Gaylord:1999}. The leaky mode is confined at the gold/air interface and has an effective index $n_{SPP1}=1.0375$ and a propagation length $L_{SPP1}=\SI{8.8}{\micro\meter}$ ($n_{SPP1}^{''}=\SI{6.07e-3}{}$). The bound SPP is confined at the gold/glass interface ($n_{SPP2}=1.663$, $n_{SPP2}^{''}=\SI{1.45e-2}{}$ corresponding to  $L_{SPP2}=\SI{3.7}{\micro\meter}$). 

Figure \ref{fig:GamaThinSlab}a presents the dipolar emitted power $P(k_{\parallel})$. We observe a similar behaviour than  above a gold mirror with three different contributions; the radiative waves ($k_{\parallel}\le n_1k_0$), the two SPP contributions and the lossy waves for high wavenumbers $k_{\parallel}$. Note that the leaky plasmon also contributes to the radiative waves via leakage into the glass substrate (so-called surface plasmon coupled emission -SPCE) \cite{Mollet-Drezet:2012}.  We again observe the strong increase of LSW for a dipolar emitter close to the metal film.  Finally, the two peaks near $k_{\parallel}=1.04k_0$ and $k_{\parallel}=1.66k_0$ reveal the SPP contributions. Their resonance profiles are shown in details on Fig. \ref{fig:GamaThinSlab}(b,c).  These two modes result from the coupling of the (leaky) Au/air and (bound) Au/glass SPP of a single interface. Therefore their resonant behaviour does not follow a Lorentzian profile anymore but rather a Fano profile \cite{Miroshnichenko:2010,Gallinet-Martin:2011}. Fano profile expresses 
 \begin{eqnarray}
P(k_{\parallel})=\frac{P(k_{spp})}{q^2+1}\frac{(x+q)^2+b}{1+x^2}\,, \text{with} \\
x=\frac{k_{\parallel}/k_0-n_{SPP}}{n_{SPP}^{''}} \,.
\end{eqnarray}
This expression is a generalization of the Fano formula to lossy materials according to ref. \cite{Gallinet-Martin:2011}. $q$ is the ratio between the optical response of the mode of interest and the second mode and $b$ introduces an offset due to losses. The Fano fits perfectly match the two SPP resonances (Fig. \ref{fig:GamaThinSlab}b). The leaky mode presents a high $q$ parameter ($q=50$) so that it closely follows a Lorentzian profile. However, the bound mode resonance cannot be fitted with a Lorentzian profile since it is coupled to a leaky mode that can be treated as a continuum, hence the Fano behaviour \cite{vanExter-Woerdman:1996}. 
\begin{figure*}
\includegraphics[width=16cm]{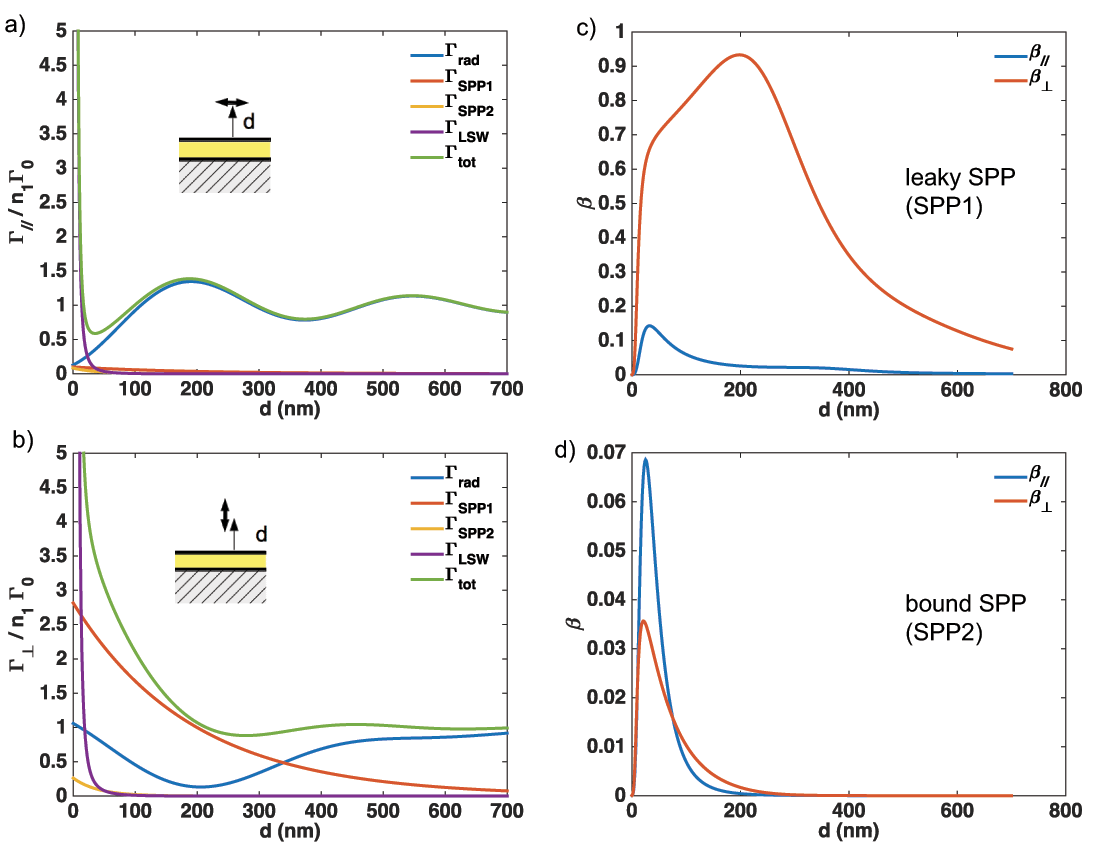}
\caption{Contributions to the total decay rate as a function of the distance to the 50 nm gold film for a dipole parallel (a) or perpendicular (b) to the surface. c,d) Coupling efficiency to the leaky (c) and bound (d) SPP. The emission wavelength is  $\lambda_{em}=\SI{670}{\nano \meter}$.}
\label{fig:GamaSlab50Contrib}
\end{figure*}
The SPP contribution is therefore estimated from the integral of the Fano resonance. If we remove the continuum background contribution, we obtain (see \ref{app:Fano})
\begin{eqnarray}
\label{eq:SPPPurcell2}
\frac{\Gamma_{SPP}}{n_1\Gamma_0}&=& \frac{p(k_{SPP})}{L_{SPP}}\frac{q^2+b-1}{q^2+b }\frac{\pi}{2},\text{with} \\
p(k_{SPP})&=&\frac{P(k_{SPP})}{n_1P_0} \,.
\nonumber
\end{eqnarray} 

We plot in Fig. \ref{fig:GamaSlab50Contrib}(a,b), the different contributions to the total decay rate for a dipolar emitter oriented parallel or perpendicular to the metal surface. {The radiative rate refers to the power integrated over $0 \le k_{\parallel} \le n_1 k_0$ (radiative waves in medium 1). Note that leakages of SPP1 into the substrate contribute to the scattering rate but not to the radiative rate. We discuss this point later.}
The contribution of surface plasmon to the decay rate remains small for a dipole parallel to the surface. Differently, for a perpendicular dipole, we observe strong excitation of the gold/air SPP1 (Fig. \ref{fig:GamaSlab50Contrib}b) whereas the gold/glass SPP2 contribution remains small due to poor overlap with the dipolar emission since the emitter is located in air. This is quantified by the coupling efficiency $\beta$ represented on Fig. \ref{fig:GamaSlab50Contrib}(c,d). Up to $\beta=93\%$ coupling efficiency is achieved for a vertical dipole $d=\SI{200}{\nano\meter}$ above the gold film. This high $\beta$-factor originates from the small radiative rate at this distance due to an interference effect between the direct emission and the reflected field. Note that the SPP rate is equal to the free-space rate at this distance ($\Gamma_{SPP1}/n_1\Gamma_0 \simeq 1$). Therefore the power coupled into the SPP guide is the same as the power emitted in the whole space ($4\pi$ str) by an isolated radiator. 

\begin{figure}
\includegraphics[width=8cm]{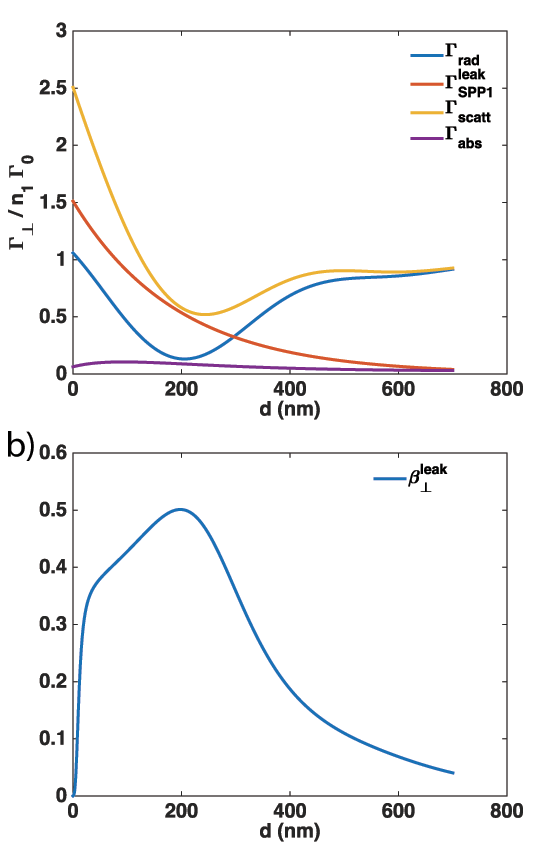}
\caption{a) Radiative and leakage contribution to the scattered decay rate. The absorption contribution is $\Gamma_{abs}=\Gamma_{rad}+\Gamma_{leak}-\Gamma_{scatt}$. b) Coupling efficiency $\beta_{leak}$ into leakage. The dipolar emitter is perpendicular to the metal surface.}
\label{fig:Slab50Leak}
\end{figure}

Since the gold/air SPP is leaky into the substrate ($n_{SPP}<n_{sub}$), it is worthwhile to estimate leakage radiation that are of interest for {\it e.g.} leakage radiation microscopy \cite{GrandidierJMicrosc:2010,Drezet2008,Bouhelier-ColasdesFrancs-Grandidier:2012} or surface enhanced fluorescence \cite{Fort-Gresillon:2005,Lakowicz5:2005}. In order to determine the leakage contribution, we first estimate the leakage and Ohmic losses of the gold/air SPP. Indeed, the finite propagation length of the leaky plasmon originates from i) intrinsic (Ohmic) losses with the rate per unit length $\alpha_i$ and ii) radiative losses into the substrate with the rate per unit length $\alpha_{leak}$ \cite{Bouhelier-ColasdesFrancs-Grandidier:2012,Raether1986}. The propagation length expresses
\begin{eqnarray}
L_{SPP}=\frac{1}{\alpha_i+\alpha_{leak}} \,.
\end{eqnarray}
The leakage rate is estimated by cancelling the ohmic losses [$Im(\varepsilon_{Au})$ put to zero] to $\alpha_{leak}=\SI{6.1e-2}{\per \micro \meter}$.
Therefore SPP leakage contributes for $\int_{0}^{\infty}\alpha_{leak}e^{-r_{\parallel}/L_{SPP}}dr_{\parallel}=\alpha_{leak}L_{SPP}=53\%$ to the SPP rate. 
Figure \ref{fig:Slab50Leak}a presents the power emitted in the far field $\Gamma_{scatt}$ as well as the radiative rate and  SPP leakage ($\Gamma_{leak}=53\%\Gamma_{SPP1}$). We also estimate the absorption in the metal $\Gamma_{abs}=\Gamma_{rad}+\Gamma_{leak}-\Gamma_{scatt}$. Only a small part of the radiative emission is absorbed in the metal and doesn't contribute to the far-field emission. This rate practically doesn't depend on the distance to the metal film as in the mirror case (see Fig. \ref{fig:GamaAbs525}). Last, the collection efficiency into the substrate using an oil immersion objective ($NA>n_{SPP1}$) is estimated as $\beta_{leak}=\Gamma_{leak}/\Gamma_{tot}$ and is shown in Fig. \ref{fig:Slab50Leak}b. It reaches $50\%$ at 200 nm.    

\subsubsection{Quality factor and mode confinement}
\label{sect:leakyQspp-Leff}
We proceed as previously (see \S \ref{sect:Qspp-Leff}) to estimate the SPPs quality factor and effective length. We gather the values achieved for the leaky and bound SPPs in table \ref{tab:leakyLeff}. Although the leaky SPP is delocalized into the substrate, we can estimate an effective length according to the Purcell factor definition. It is again close to the penetration depth in air, indicating that the confinement entering the Purcell factor corresponds to the near-field behaviour of the SPP mode. We will observe a similar behaviour for localized surface plasmons (see \S \ref{sect:Mie}).
\begin{table}
\begin{tabular}{cccccc}
&$Q$& $L_{eff}$ (nm) & $\delta/2$ (nm) &$n_g$ & $F_p$ \\
\hline
leaky  & 85 &159  &193 &0.92& 3.02 \\
bound  & 57 & 27  &76 &0.57&5.1
\end{tabular}
\caption{Quality factor and effective length of leaky and bound SPPs. 
The penetration depth  (in air or glass) is indicated for comparison. The group index and Purcell factor [SPP decay rate calculated at the gold/air (leaky SPP) and gold/glass (bound SPP) interfaces] used to estimate $L_{eff}$ are also given.}
\label{tab:leakyLeff}
\end{table}

\subsection{In-plane plasmonic cavity}
\begin{figure}
\includegraphics[width=8cm]{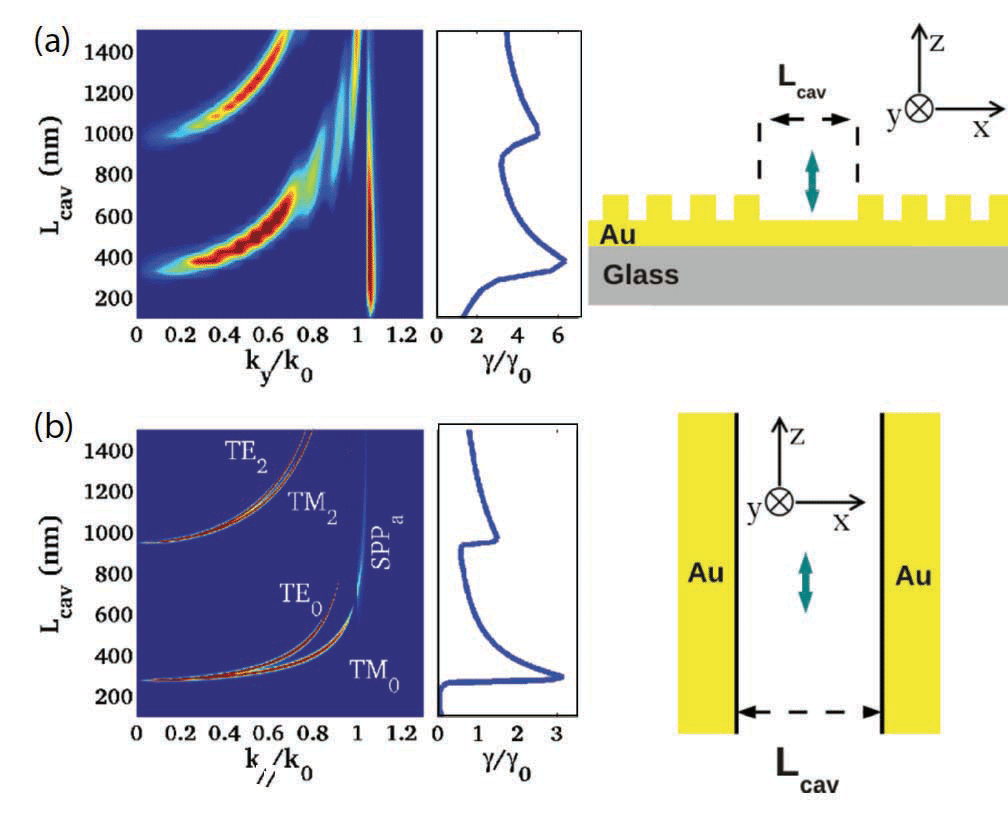}
\caption{a) Power dissipated by a dipolar emitter
inside the plasmonic cavity, as a function of cavity length $L_{cav}$ and
in plane wave vector along the cavity axis $k_y/k_0$. The glass/gold/air
slab contribution is subtracted to characterize the cavity effect. Inset:
Decay rate as a function of the cavity length obtained by integrating
the dissipated power over all the $k_y/k_0$ spectrum range (including the
glass/gold/air slab contribution). b) Same as (a) for a 1D gold/air/gold
cavity. The dipolar emitter is located at the cavity center and parallel to
the mirror walls. The 1D cavity modes are indicated on the dispersion
curve. Reproduced with permission from ref. \cite{Derom-Hermier-GCF:2014}, APS, copyright 2012.}
\label{fig:cavity}
\end{figure}

SPP is a surface wave, intrinsically confined near the metal film. It can be furthermore laterally confined by distributed Bragg reflectors, forming an in-plane plasmonic cavity as schemed in Fig. \ref{fig:cavity} \cite{NanoLettWeeberGCF:2007,Gonga-Vukovic:2007,Derom-Hermier-GCF:2014}.  Such an open in-plane plasmonic cavity would further increase the plasmonic Purcell factor, with the possibility to access the fluorescent emitter. This enables an external control or manipulation of the emitter position (optical trapping, AFM manipulation, ...) or emission properties (Stark effect using a STM tip, ...). 

The two-dimensional dipolar emission can be numerically achieved using the 2D-Green's dyad technique. It expresses \cite{Barthes-GCF-Bouhelier-Weeber-Dereux:2011,Barthes-Bouhelier-Dereux-GCF:2013}
\begin{eqnarray}
\frac{\Gamma_u(x,z)}{n_1\Gamma_0}=\frac{6}{n_1k_0}\int_{0}^{+\infty}ImG^{2D}_{uu}(\mathbf{r_{xz}},\mathbf{r_{xz}},k_y) dk_y 
\label{eq:gamma2D}
\end{eqnarray}
where $\mathbf{r_{xz}}=(x,z)$ is the emitter position in the transverse plane $(Oxz)$ and $k_y$ the component of the wavector along the invariant $y-$axis. Since the 2D-Green's dyad ${\bf G^{2D}}$ associated to the grating structure can be numerically computed, we can define the normalized  dipolar emission power $\frac{P(k_y)}{n_1P_0}= 
\frac{6}{n_1k_0}ImG^{2D}_{uu}(\mathbf{r_{xz}},\mathbf{r_{xz}},k_y)$ as a function of the propagation constant  $k_y$. This makes a direct analogy between the dipolar emission in 1D and 2D geometries and all the above discussion near a flat metal film is easily extended to this configuration.  

Figure \ref{fig:cavity} compares the behaviour of a (2D) planar plasmonic cavity and a (1D) Fabry-Perot cavity. The 2D dipolar emission 
$P(k_y)$ and 1D dipolar emission $P(k_\parallel)$ are calculated for a dipolar emitter parallel to the mirrors and located at the center of the cavity. The dipolar emission significantly increases for cavity size $L_{cav}=(2p+1)\lambda_{eff}/2$ with $p$ an integer and 
$\lambda_{eff}$ the mode effective wavelength ($\lambda_{SPP}$ or $\lambda_{em}/n_1$ in the plasmonic and Fabry-Perot cavities, respectively). This corresponds to the emission into the even modes of the cavity that presents an antinode at the cavity center. We observe a cut-off for cavity size below $\lambda_{eff}/2$.  The normalized decay rate, calculated as a function of the cavity size, presents very similar behaviour for these two cavities. This demonstrates the strong analogy between the in-plane plasmonic cavity and the micro-optical cavity. However, some distinct features appear in the planar cavity. First, since SPP are polarized perpendicular to the metal film, we do not observe polarization degeneracy (TE/TM) in the planar cavity. We also note the permanent contribution of the planar SPP mode at $k_y/k_0=n_{SPP}=1.04$. The gold/glass SPP at $k_y/k_0=1.66$ contribution is very weak (not shown). 
Finally, a planar plasmonic cavity relies on the confinement of the SPP surface waves instead of confining a bulk mode for an optical micro-cavity.  This additionnal mode confinement leads to a significant increase of the decay rate inside the cavity \cite{Derom-Hermier-GCF:2014}. This is however at the price of strong losses.  

\section{Purcell factor near a plasmonic waveguide}
\label{sect:SPPguide}
As discussed above, SPP Purcell factor into extended metal film can be increased by an in-plane cavity that confines laterally the delocalized SPP. However, the lateral confinement is still limited to about $\lambda_{SPP}/2$.
Stronger lateral mode confinement can be achieved in plasmonic waveguides. For instance, metal nanowires does not present a cut-off and the lateral mode confinement is not diffraction limited but given by the nanowire cross-section \cite{Takahara:1997}. 
Therefore, we expect high coupling efficiency of a dipolar emitter to a metal nanowire \cite{Chang-Sorensen-Hemmer-Lukin:2007,Kolesov2009,Huck-Andersen:2009,Gruber-Krenn:2012,Ropp-Waks:2013,Bozhelvonyi-Quidant:2015,Norris:2015}. Metal nanowires define 1D plasmonic waveguides with a great potential for integrated optical routing
\cite{Wei-Xu:2012}. They can be chemically synthesized with high crystallinity hence
supporting SPP with reduced losses \cite{Ditlbacher2005,Laroche-Vial-Roussey:2007,Song-Dujardin-Zhang-GCF:2011,Viarbitskaya:2013,
Viarbitskaya-Dujardin:2013}. 
SPP propagation can be controlled by a single emitter, leading to the concept of single photon transistor \cite{Chang-Sorensen-Demler-Lukin:2007}. 
Reciprocally, two distant emitters can be interfaced via surface plasmons, with applications such as SPP mediated resonant energy transfer \cite{MartinCano-MartinMoreno-GarciaVidal-Moreno:2010,Barthes-Bouhelier-Dereux-GCF:2013}, remote qubits entanglement \cite{GonzalezTuleda-GarciaVidal:2011,Chen-Nori:2011} 
or nano-optical logical gates \cite{Fleischhauer:2010}. SPP gain  amplification has also been shown when dipolar emitters play the role of gain medium \cite{GrandidierNanoLet:2009,KenaCohen-Maier:2013,Paul-Nordlander-Link:2014}.

On the basis of expression (\ref{eq:gamma2D}) the different decay channels near a plasmonic waveguide are easily estimated. In order to specifically  discuss the contributions of the guided modes, we rewrite the expression (\ref{eq:gamma2D}) in the equivalent form 

\begin{eqnarray}
\frac{\Gamma(x,z)}{n_1\Gamma_0}=1+\frac{3\pi}{n_1^3k_0}\int_{0}^{+\infty}\frac{\Delta \rho ^{2D}_{u}(\mathbf{r_{xz}},k_y)}{k_y} dk_y 
\label{eq:gamma2DRho}
\end{eqnarray}
where we have introduced the 2D-LDOS \cite{Barthes-GCF-Bouhelier-Weeber-Dereux:2011,Barthes-Bouhelier-Dereux-GCF:2013} 
\begin{equation}
\Delta\rho^{2D}(\mathbf r_{xz},k_y)=\frac{2\varepsilon_1 k_y}{\pi} Im  
\Delta  G^{2D}_{uu} (\mathbf r_{xz},\mathbf r_{xz},k_y) 
\end{equation}
and 
$\Delta \mathbf G^{2D}=\mathbf G^{2D}-\mathbf G^{2D}_0$ is the difference between the total Green's tensor of the 2D-structure and the free-space Green's tensor $\mathbf G^{2D}_0$. It describes the role of the waveguiding structure only. 
\begin{figure}
\includegraphics[width=8cm]{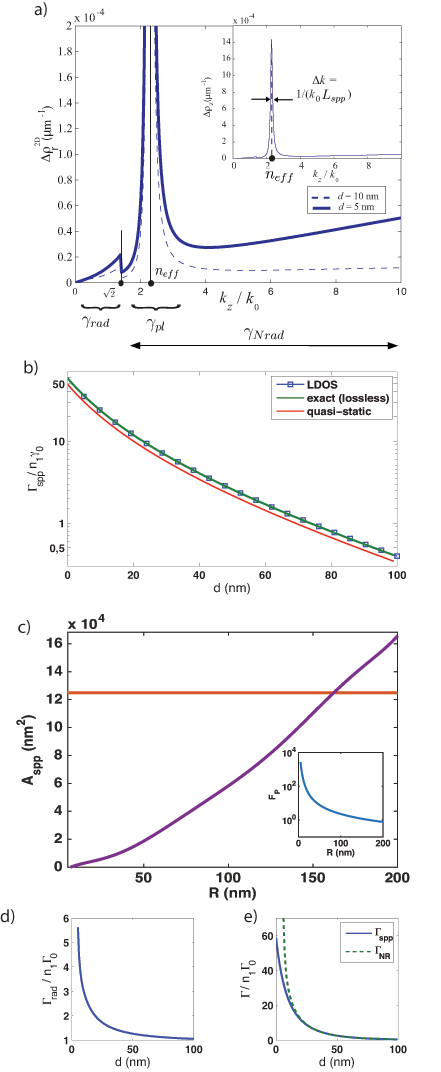}
\caption{a) Radial 2D-LDOS near a circular silver nanowire as a function of $k_y$ for two distances $d$ to the wire surface. b) Comparison of SPP rate into the plasmonic wire for a real lossy waveguide (LDOS), an ideal lossless waveguide and within the quasi-static approximation (in presence of losses). c) Mode effective area of circular waveguide as a function of its radius. The horizontal line indicates the diffraction limit $(\lambda/2n_1)^2$ [inset: Purcell factor (Eq. \ref{eq:FP2DId}). d) Radiative rate near a circular nanowire ($R=20$ nm). e) Non radiative rate $\Gamma_{NR}$ and SPP contribution $\Gamma_{SPP}$ as a function of distance to the nanowire surface. The wire radius is $R=20$ nm (except in c) and the emission wavelength is $\lambda_{em}=\SI{1}{\micro\meter}$. The permittivity of the wire and surroundings permittivity are $\varepsilon=-50+3.85i$ and $\varepsilon_1=2$, respectively. Adapted with permission from ref. \cite{Barthes-GCF-Bouhelier-Weeber-Dereux:2011}, APS, copyright 2011.}
\label{fig:SPPcirc}
\end{figure}

Figure \ref{fig:SPPcirc}a represents the variation of the 2D-LDOS near a silver nanowire as a function of the propagation constant $k_y$. It behaves very similarly to the mirror case (Fig. \ref{fig:PSlab}). In particular, we can again distinguish three contributions to the decay rate. i) Radiative waves for $k_y<n_1k_0$, ii) SPP peaked near the plasmon propagation constant (here $k_{SPP}=2.28k_0$) and iii) lossy waves at large $k_y$. The corresponding decay channels are calculated by numerical integrations over the different wave domains. 

We first discuss the SPP rate. The SPP contribution follows a Lorentzian profile as above a thick plasmon film. It is peaked on the plasmon propagation constant $k_{SPP}=2.28k_0$ and with a FWHM inversely proportional to the propagation length ($L_{SPP}=\SI{1.2}{\micro\meter}$; see the inset of figure \ref{fig:SPPcirc}a).
This again leads to a closed form expression of the plasmonic Purcell factor \cite{Barthes-GCF-Bouhelier-Weeber-Dereux:2011,Barthes-Bouhelier-Dereux-GCF:2013}
\begin{eqnarray}
\label{eq:SPPPurcell2}
\frac{\Gamma_{SPP}}{n_1\Gamma_0}&=&\frac{3\pi\lambda}{4n_1^3k_{SPP}}\frac{\Delta \rho ^{2D}_{u}(\mathbf{r_{xz}},k_{SPP})}{L_{SPP}} \\
&=&\frac{\pi}{2}\frac{p(k_{SPP})}{L_{SPP}},\text{with} ~~ p(k_{SPP})=\frac{P(k_{SPP})}{n_1P_0} \,.
\nonumber
\end{eqnarray}

It is worth to compare this expression to the lossless case for which the coupling rate to the guided mode expresses \cite{Snyder-Love:1983,Chen-Nielsen-Gregersen-Lodahl-Mork:2010}
\begin{eqnarray}
 \frac{\Gamma_{SPP}}{n_1\Gamma_0}&=&=\frac{3\pi \epsilon_0 c \vert E_u (d)\vert ^2}{n_1k_0^2\int_{\mathbf A_\infty}(\mathbf{E}\wedge\mathbf{H^\star})\cdot  \hat \mathbf y d\mathbf A}
 \label{eq:PurcellNoLoss}
\end{eqnarray}
 where $(\mathbf{E},\mathbf{H})$ is the electromagnetic field associated with the guided SPP mode. For a circular waveguide, an analytical expression is readily obtained (see supplementary information of ref. \cite{Barthes-Bouhelier-Dereux-GCF:2013}). Figure \ref{fig:SPPcirc}b compares the SPP rate obtained using the 2D-LDOS formalism in the lossy nanowire (eq. \ref{eq:PurcellLoss}) and the lossless ideal system (eq. \ref{eq:PurcellNoLoss}). Lossy and lossless rates perfectly superimpose, revealing that the coupling rate to the guided mode does not depend on the propagation losses \cite{Barthes-GCF-Bouhelier-Weeber-Dereux:2011}.

The equivalence between the lossy and lossless Purcell factors for a guided mode can be understood as follows. The dispersion relation $\omega=f(k_{spp})$ governs the guided mode excited by an emitter near the emission angular frequency $\omega_{em}$. Therefore, the dirac distribution in the decay rate Eq. (\ref{eq:GammaDirac}) becomes 
\begin{eqnarray}
\delta(\omega-\omega_c) = \frac{1}{v_g}\delta[k_{spp}-k_{spp}(\omega_{em})]
\end{eqnarray}
introducing the group velocity $v_g=\partial \omega/\partial k_{spp}$ and assuming a weak dispersion around the emission wavelength. The dirac distribution $\delta[k_{spp}-k_{spp}(\omega_{em})]$ is again replaced by a Lorenzian profile for a lossy waveguide so that we achieve 
\begin{eqnarray}
F_p &=&\frac{3}{4 \pi^2}  \left ( \frac{\lambda_{em}}{n_1} \right )^{3} ~ \frac{Q_{spp}}{V} \frac{\omega_{em}}{v_g k_{spp}} \,.
\end{eqnarray}
Moreover, the quality factor and effective volume of the guided mode are expressed, respectively by 
\begin{eqnarray}
Q_{spp}&=&\frac{k_{guide}}{\Delta k_{guide}}=k_{spp}L_{spp} \;
\end{eqnarray}
and
\begin{eqnarray}
V&=&\frac{\int \varepsilon({\bf r}) \vert \mathbf{ { E}}(\mathbf{ r}) \vert^2  dxdydz }{Max\left[\varepsilon_1\vert \mathbf{ { E}}(\mathbf{ r}) \vert^2\right]} \\
\nonumber
&=&\frac{\int \varepsilon(x,z) \vert \mathbf{ { E}}(x,z) \vert^2  dxdz}{Max\left[\varepsilon_1\vert \mathbf{ { E}}(x,z) \vert^2\right]} \times \int_{-\infty}^\infty e^{-\vert y\vert /L_{spp}}  dy \\
&=&A_{eff}2L_{spp}\;.
\end{eqnarray}
$A_{eff}$ defines the mode effective area (see Eq. \ref{eq:Amode}) and we assume again a non dispersive medium. Finally, the Purcell factor simplifies to the nanofiber expression (\ref{eq:Purcell2D})
\begin{eqnarray}
F_p=\frac{\Gamma_{spp}}{n_1\Gamma_0}=\frac{3}{4 \pi}  \frac{(\lambda_{em}/n_1)^2}{A_{eff}} \frac{n_g}{n_1}
\end{eqnarray}
and does not depend on the propagation length. Moreover, the guided SPP mode volume obeys $V_{spp}=2A_{eff}L_{spp}$ that becomes strongly confined for short propagation distances $L_{spp}$. It makes a bridge between delocalized and localized SPP, notably in the quasi-static regime where they present very similar behaviour \cite{Barthes-Bouhelier-Dereux-GCF:2013}. This discussion also points out that the role of losses in the plasmonic Purcell factor must be carried out carefully.  

Eventually, the Purcell factor is identical considering a real lossy or an ideal non-lossy plasmonics waveguide, so that we can identify the Purcell factor expressions (\ref{eq:Purcell2D}) and (\ref{eq:PurcellNoLoss}) to estimate the SPP confinement. As previously, we identify at the nanowire surface, 
\begin{eqnarray}
\label{eq:FP2DId}
&&\frac{\Gamma_x}{n_1\Gamma_0}+\frac{\Gamma_y}{n_1\Gamma_0}+\frac{\Gamma_z}{n_1\Gamma_0}=F_p \,,\\
&&\equiv\frac{3}{4 \pi}  \frac{(\lambda_{em}/n_1)^2}{A_{eff}} \frac{n_g}{n_1} \,.
\label{eq:Purcell2DIdent}
\end{eqnarray} 

The effective surface of the guided SPP is shown on Fig. \ref{fig:SPPcirc}c as a function of the wire radius. This reveals that quantum plasmonics relies on strongly subwavelength mode confinement with huge Purcell factor (up to $10^3$ see inset of Fig. \ref{fig:SPPcirc}c) \cite{Chang-Sorensen-Hemmer-Lukin:2006,Russel-Yeung-Hu:2012}. We achieve similar values for $\lambda_{em}=\SI{670}{\nano \meter}$ and a gold nanowire (not shown). The small SPP effective area permits to decrease the threshold for SPP amplification compared to a photonic nanowire of similar cross-section \cite{Oulton-Zhang:2009}. 
\begin{figure}
\includegraphics[width=8cm]{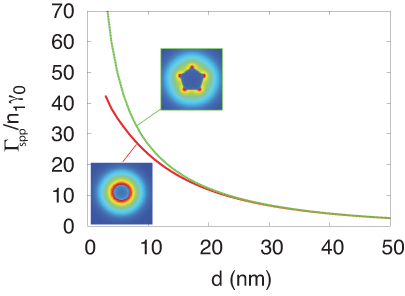}
\caption{Purcell factor near a circular or pentagonal silver nanowire ($\lambda_{em}=\SI{1}{\micro \meter}$). $d$ is the distance to the nanowire surface or edge. Reproduced with permission from ref. \cite{Barthes-GCF-Bouhelier-Weeber-Dereux:2011}, APS, copyright 2011.}
\label{fig:Aspp}
\end{figure}

Finally, the radiative and non radiative contributions are presented in Fig. \ref{fig:SPPcirc}(d,e). Above $d=20$ nm, the only contribution to the non radiative rate is SPP, due to losses along the propagation. At short distances, electron scattering are responsible of the additionnal losses, leading to fluorescence quenching. Last, a coupling efficiency into the guided plasmon of $\beta=83\%$ is achieved at 20 nm from the nanowire (not shown) \cite{Barthes-GCF-Bouhelier-Weeber-Dereux:2011}.  

For a nanowire above a glass substrate, the guided SPP becomes leaky. The 2D-LDOS again follows a Fano profile (or a Lorentzian profile for weak leakage) and the leakage rate is easily estimated as done in the case of thin metal film. For instance, $\beta_{leak}=70\%$ of the emission is collected into the substrate for a 100 nm silver wire 50 nm above the glass substrate  \cite{Barthes-GCF-Bouhelier-Weeber-Dereux:2011}.

Since the 2D-Green's dyad associated to a plasmonic waveguide can be numerically computed, this formalism can be applied to arbitrary geometries  \cite{ColasdesFrancsPRB:2009,Paulus2001,Kottmann-Martin-Smith-Schultz:2001,ColasdesFrancs-Hugonin:2011}. Specifically, crystalline silver nanowires present a pentagonal cross-section shape leading to strong mode confinement at the corners (see Fig. \ref{fig:Aspp}) \cite{Song-Dujardin-Zhang-GCF:2011,Nauert-Nordlander-Link:2014}. This mode confinement significantly increases the Purcell factor near the edge of the nanowire compared to a circular nanowire \cite{Barthes-GCF-Bouhelier-Weeber-Dereux:2011}. 

\section{Localized plasmon}
\label{sect:LSP}
This last section is devoted to localized plasmon for which we expect full 3D subwavelength confinement. For the sake of clarity, we focus on a spherical metal nanoparticle (MNP) that constitutes a canonical configuration for LSP. We first consider the quasi-static approximation for which we derive analytical expressions for the mode volume and quality factor of each mode (\S \ref{sect:QS}). In section \ref{sect:Mie}, we extend the discussion to the retarded regime using Mie theory. 

\subsection{Quasi-static regime}
\label{sect:QS}
\subsubsection{Dipolar LSP}
Let us consider a spherical metal nanoparticle (MNP) of radius $R$ small compared to the wavelength. For clarity, we first discuss the dipolar response of the particle.  It is characterized by the effective polarizability 

\begin{eqnarray}
\alpha_1^{eff}=\left[ 1 - i~\frac{2k^{3}}{3}\alpha_1 \right] ^{-1} \alpha_1 \;\;, (k=2\pi / \lambda) \,, \\
\alpha_1(\omega)=\frac{\varepsilon_m(\omega)-1}{\varepsilon_m(\omega)+2} R^3 \,,
\label{eq:alpha_eff1}
\end{eqnarray}
The so-called radiative reaction ($2k^3 \alpha_1/3$) originates from finite size effect and energy conservation considerations \cite{Jackson:1998,GCFIJMS:2009,LeRu-Auguie:2013,Grigoriev-Stout:2015}.
$\alpha_1$ is the nanoparticle quasi-static (dipolar) polarisability and $\varepsilon_m$ is the metal dielectric constant. The dipole plasmon resonance appears at $\omega_1$ such that $\varepsilon_m(\omega_1)+2=0$. In case of Drude metal, the dipolar resonance is $\omega_1=\omega_p/\sqrt{3}$ with $\omega_p$ the bulk metal plasma angular frequency.

If the metal dielectric constant obeys a Drude model, the effective polarizability  follows a Lorentzian profile near the LSP resonance \cite{Carminati-Greffet-Henkel-Vigoureux:2006}  
\begin{eqnarray}
\label{eq:alpha1Drude}
\varepsilon_m&=&1-\frac{\omega_p^2}{\omega^2+i\kappa_{abs} \omega} \,;\\
\alpha_1^{eff}(\omega)&\underset{\omega_1}{\sim}&\frac{\omega_1}{2(\omega_1-\omega)-i\kappa_1} R^{3}\,, \\
\kappa_1&=&\kappa_{abs}+\frac{2(k_1 R)^3 \omega_1}{3}  \,, (k_1=\omega_1/c) \,.
\end{eqnarray}
$\kappa_1$ is the decay rate of the particle dipolar mode, and includes both the Joule ($\kappa_{abs}$) and radiative [$\kappa_1^{rad}=2(k_1R)^3 \omega_1/3$] losses rates. We can therefore define the quality factor of the dipolar mode 
$Q_{1}=\omega_1/\kappa_1$ that typically ranges from 10 to 25 for gold or silver nanoparticles \cite{GCFIJMS:2009}.  

The decay rate of a dipolar emitter located in the very near-field of a spherical metal particle approximates to \cite{JCPGCF:2005,Carminati-Greffet-Henkel-Vigoureux:2006,OptExpGCF:2008}
 \begin{eqnarray}
 \frac{\Gamma_{1}^{\perp}}{\Gamma_0}&\sim& \frac{6}{k^3 z_0^6}
 Im(\alpha_1) \,, \\
 \label{eq:gz}
 &\underset{\omega_1}{\sim}&\frac{6 \omega_1 R^3}{k_1^3 z_0^6 \kappa_1}\sim \frac{3}{4\pi^2}\lambda_1 ^3
\frac{R^3}{\pi z_0^6} Q_{1} \,  \\
  \frac{\Gamma_{1}^{\parallel}}{\Gamma_0}&\sim& \frac{3}{2k^3 z_0^6}
 Im(\alpha_1) \underset{\omega_1}\sim \frac{3}{4\pi^2}\lambda_1 ^3
\frac{R^3}{4\pi z_0^6} Q_{1} \, 
 \label{eq:gx}
 \end{eqnarray} 
for a dipole emitter oriented perpendicular or parallel to the nanoparticle surface and an emission tuned to the dipolar particle resonance ($\lambda_{em}=\lambda_1=2\pi c/\omega_1$). 
In order to determine the dipolar mode effective volume, we now identify the coupling rate $\Gamma/\Gamma_0$ to the Purcell factor (Eq. \ref{eq:Purcell}). It is worthwile to note that the Purcell factor is obtained assuming a single mode cavity. Since the dipolar mode is three-fold degenerated, the SPP rate writes $\Gamma_{1}/\Gamma_0=3F_p$ so that we can identify the Purcell factor to the SPP rate of a randomly oriented dipolar emitter at the particle surface ($z_0=R$). The average decay rate writes
\begin{eqnarray}
\frac{\langle \Gamma_{1}\rangle}{\Gamma_0}=\frac{\Gamma_{1}^{\perp}+2\Gamma_{1}^{\parallel}}{3\Gamma_0}\sim \frac{3}{4\pi^2}\lambda_1 ^3
\frac{1}{2\pi R^3} Q_{1}=\frac{3}{4\pi^2}\lambda_1 ^3
\frac{Q_{1}}{V_1}
\end{eqnarray}
where we define the dipolar LSP mode volume $V_1=2\pi R^3=3/2~V_{0}$. $V_0=4\pi R^3/3$ is the volume of the MNP.
As a consequence, although the dipolar LSP is radiative in the far-field, it is possible to assign a finite mode volume within the quasi-static approximation. As expected, it is strongly subwavelength for small MNP \cite{Greffet-Laroche-Marquier:2010,GCFIJMS:2009,Deeb-Bachelot-Plain-Soppera:2010,Zhou-Plain:2014}.

\subsubsection{High order modes}
The Purcell factor associated to the dipolar mode can be generalized to each LSP mode. {Due to the (2n+1) degeneracy, the coupling rate to the $n^{th}$ mode (n=1, dipolar LSP, n=2, quadrupolar LSP, ...) obeys \footnote{We use a different definition for the mode volume in Ref. \cite{GCFIJMS:2009}, assuming a randomly oriented dipole. The definition used here, that includes  the mode degeneracy is more consistent with usual Purcell factor definition.}}
\begin{eqnarray}
\frac{\Gamma_{n}}{\Gamma_0}&=&(2n+1) F_p\,, \text{with}\\
F_p&=&\frac{3}{4\pi^2}\lambda_1 ^3 \frac{Q_{n}}{V_n}\,, \\
Q_n&=&\frac{\omega_n}{\kappa_n}=\frac{\omega_n}{\kappa_{abs}+\kappa_n^{rad}} \,, \label{eq:QnLSP}\\
\label{eq:VnQS}
{V_n}&=&\frac{3}{n+1}V_0 \,,\\
\kappa_n^{rad}&=&\omega_n \frac{(n+1)(k_n R)^{2n+1}}{n(2n-1)!!(2n+1)!!} \,, (k_n=\omega_n/c) \,.
\end{eqnarray}
The quality factor is governed by ohmic losses $\kappa_{abs}$ in the MNP, identical for all the modes, and radiative losses that slightly decreases for high order modes [$\kappa_n^{rad}\propto (k_n R)^{2n+1}$]. Quality factor therefore slightly increases from a few 10 for a dipolar LSP ($n=1$) to about 50 to 100 for high order modes. In addition, the high order modes are more confined than the dipolar LSP ($V_1=1.5V_0$, $V_2=V_0$, $V_n \underset{n\rightarrow \infty}\rightarrow 0$). This quantifies the mode confinement and reveals that although having low Q factor, LSP presents sub diffraction mode volume that ensures efficient coupling to a nearby emitter at the origin of surface enhanced spectroscopies. 

Last, we can again define a coupling efficiency $\beta$-factor to a given mode. It strongly depends on the distance to the MNP but also of the emission wavelength. For an emission wavelength matching the dipolar LSP, the coupling efficiency reaches $\beta_1=90\%$ into the n=1 mode at a distance $d=10$ nm. It can be as high as  $\beta_2=87\%$ into the quadrupolar mode at $d=15$ nm if the wavelength emission matches the n=2 LSP resonance (not shown, see ref. \cite{GCFIJMS:2009}).

\subsubsection{{Energy confinement}}
\begin{figure}
\includegraphics[width=8cm]{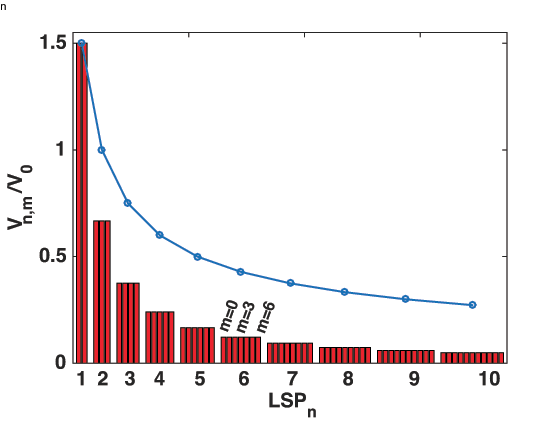}
\caption{Energy confinement mode volume $V_{l,m}$ as a function of the mode number $l$ and $m=0,1,\ldots l$. The curve is the Purcell volume estimated identifying the decay rate to a given mode and Purcell factor $V_{Purcell}=3V_0/(l+1)$.}
\label{fig:Vlm}
\end{figure}

In cQED, the mode volume is defined as the energy confinement (see eq. \ref{eq:Vmode}).   For dispersive materials, this \emph{extrapolates} to
\begin{eqnarray}
\label{eq:Vnrj}
V_n^{nrj}=\frac{\int  U_n({\bf r}) d{\bf r}}{max[\varepsilon_0\varepsilon_1~|{\bf E_n}({\bf r})|^2]} \;, \\
U_n({\bf r})=\frac{\partial[\omega\varepsilon_0\varepsilon({\bf r},\omega)]}{\partial \omega} ~|{\bf E_n}({\bf r})|^2
+\mu_0|{\bf H_n}({\bf r})|^2
\nonumber
\end{eqnarray}
where (${\bf E_n},{\bf H_n}$) is the electromagnetic field of the $n^{th}$ mode. In the quasi-static regime, the magnetic contribution is negligible and the electric field is confined near the particle surface. This definition then leads to $V_n^{nrj}=6/(n+1)^2V_0$ \cite{Khurgin-Sun:2009}. 
The effective volumes of the (2n+1) degenerated $n^{th}$ LSP modes are represented on Fig. \ref{fig:Vlm} (see \ref{sect:LSPVnrj} for details). We observe that the (2n+1) degenerated modes present the same volume (note that the energy confinement of the (l,m) mode is normalized with respect to the (l,m=0) mode maximum intensity).

The ratio between the mode volume deduced from Purcell factor ($V_{Purcell}$) and the mode volume estimated from energy mode confinement assuming a lossless metal ( $V_{energy}$) 
\begin{eqnarray}
\frac{V_{Purcell}}{V_{energy}}=\frac{n+1}{2}
\end{eqnarray}
depends on the mode number $n$ only. We attribute this difference to the (Joule and radiative) losses that are neglected in the energy confinement but taken into account in the Purcell volume derivation. Unlike delocalized SPP for which Purcell factor does not depend on Joule losses, Purcell factor for a localized LSP is strongly affected by losses in the metal.  
In addition, we observe that both Purcell and energy confinement derivations leads to identical values for the dipolar mode volume. This again shows that the mode volume entering in the Purcell factor is governed by the mode near-field behaviour, even for a leaky mode (see also \S \ref{sect:leakyQspp-Leff}). In addition, it is worth noticing that the LSP volume derived from the Purcell factor, taking into account losses, does not depends on the losses in the quasi-static approximation. As for cQED, the effect of losses on the Purcell factor are fully included in the quality factor (Eq. \ref{eq:QnLSP}). 

\subsection{Retarded regime} 
\label{sect:Mie}
\begin{figure}
\includegraphics[width=8cm]{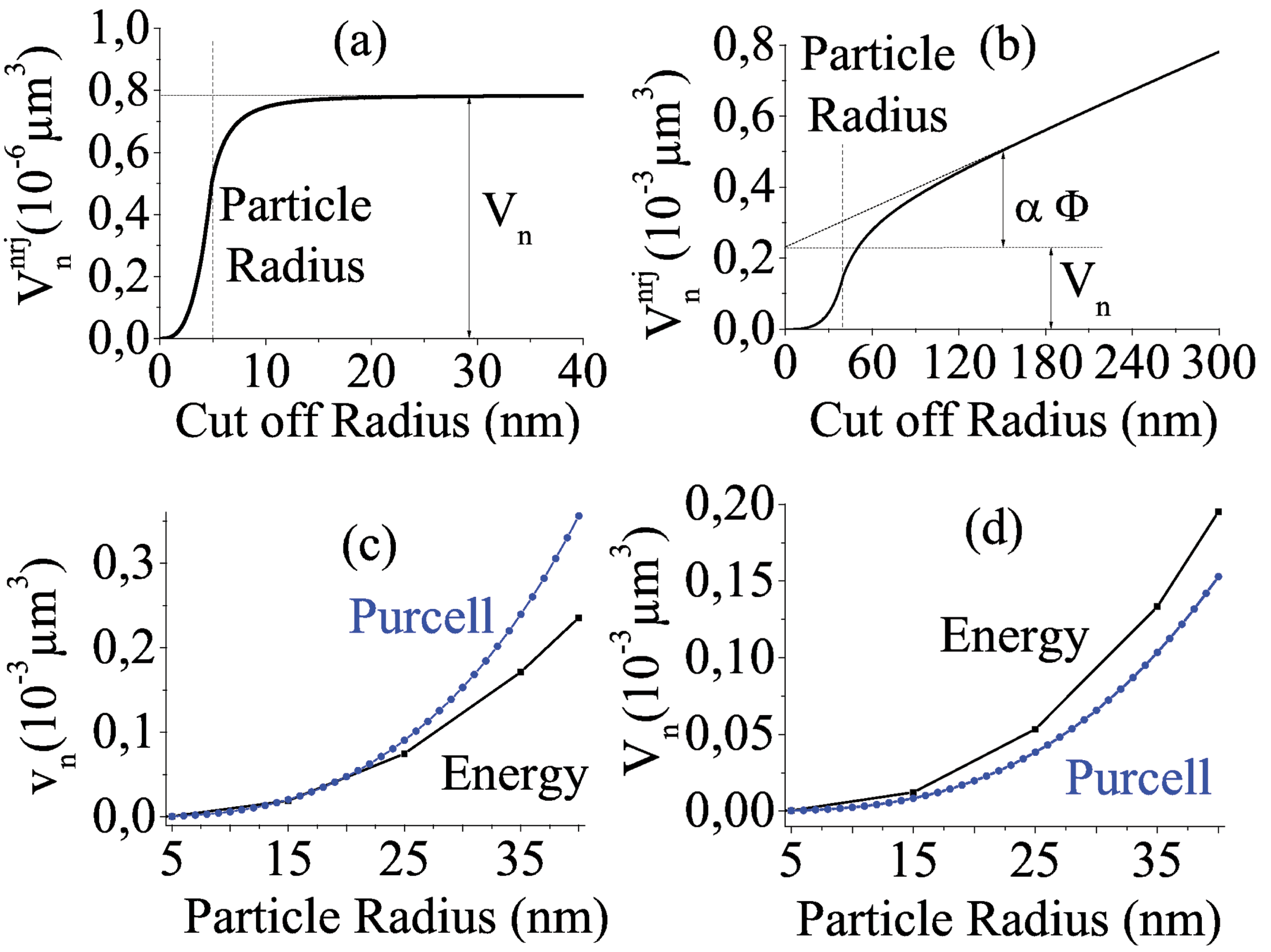}
\caption{Mode energy confinement. (a,b) Mode volume  (Eq. \ref{eq:Vnrj}) as a function of the integration sphere cut-off radius. 
The silver sphere radius is $a=5~nm$ (a, dipolar resonance at $\lambda_{1}=\SI{325}{\nano \meter}$) or $a=40~nm$ (b, dipolar resonance at $\lambda_{1}=\SI{365}{\nano \meter}$, quadrupolar resonance at  $\lambda_{2}=\SI{320}{\nano \meter}$). The energy enclosed in the integration sphere linearly increases with sphere 
radius with a slope proportional to the field flux $\Phi$. The mode volume energy is defined as the ordinate at origin. 
(c,d) Dipolar and quadrupolar LSP volumes (the wavelength depends on the particle size and considered mode). Purcell and energy refer to the mode volume definition. 
The metal dielectric constant follows Drude model. Reproduced with permission from ref. \cite{Derom-Vincent-Bouhelier-GCF:2012}, EPL, copyright 2012.}
\label{fig:Vnrj}
\end{figure}

So far, we have discussed the Purcell factor within the quasi-static regime. In this section, we generalize it to spherical MNP of arbitrary size using Mie expansion. Particular attention is again devoted to the definition of the mode volume as the energy confinement, in analogy with cQED definition.
Dipolar emission near a spherical particle of arbitrary size is exactly solved using Mie (modal) expansion \cite{Kim-Leung-George:1988}.
We can therefore again define a mode volume by identifying the contribution to the total decay rate of the $n^{th}$ mode to the Purcell expression (\ref{eq:Purcell})\cite{GCF-Derom-Vincent-Bouhelier-Dereux:2012}. In the following, we define this volume as the Purcell effective volume. We compare it to the mode energy confinement (see eq. \ref{eq:Vnrj}).   
However, the application of definition (\ref{eq:Vnrj}) is difficult in the retarded regime since LSPs leak in the far-field. 
As far as light-matter coupling is concerned, the pertinent parameter is the confinement of the mode energy stored 
inside the cavity. Following the work of Koenderink \cite{Koenderink:2010}, the intrinsic mode volume is 
estimated from the energy confinement, excluding radiative leaks [i.e, ordinate at the origin in Fig. \ref{fig:Vnrj}(a,b)].

In figure \ref{fig:Vnrj}, we compare the mode volume estimated from Purcell factor and energy definition as a function of the MNP size. This again reveals sub diffraction mode volume ensuring efficient light-matter  interaction. For small particle, we recover the quasi-static limit. For large particles, the energy definition is in qualitative agreement with the Purcell factor definition. However, large MNP supports quasi-mode with lifetime shorter than the collective oscillation so that the concept of Purcell factor fails to describe this regime.

\subsection{Generalisation of the Purcell factor concept}
\subsubsection{Complex mode volume}
Recently, Sauvan and coworkers define a complex mode volume \cite{Sauvan-Lalanne:2013}
\begin{eqnarray}
\label{eq:VnrjPML}
V_n=\frac{\frac{1}{2}\int \bf \tilde E_n \cdot \frac{\partial (\omega\varepsilon_0\varepsilon)}{\partial \omega} \cdot {\bf \tilde E_n}-\mu_0{\bf \tilde H_n}^2 d{\bf r}}{max[\varepsilon_0\varepsilon_1~[{\bf \tilde E_n}({\bf r})]^2]} 
\end{eqnarray}
where (${\bf \tilde E_n},{\bf \tilde H_n}$) is the quasi-normal $n^{th}$ mode. The convergency of the volume integral is ensured thanks to phase matching layers.  This generalises the Purcell factor to dissipative cavities such that 
\begin{equation}
F_p = \frac{\Gamma_n}{n_1~\Gamma_0} = \frac{3}{4 \pi^2}  \left ( \frac{\lambda}{n_1} \right )^{3} ~Re \left(\frac{Q_n}{V_n}\right) \;. 
\label{eq:PurcellLoss}
\end{equation}
This definition is consistent with the Purcell volume definition (Fig. \ref{fig:Vnrj}) and reconciles the Purcell factor with the energy-like confinement definition of the mode volume \cite{ZambranaPuyalto-Bonod:2015,Kristensen-Ge-Hughes:2015}. Similar definitions should work for the effective mode area or length. 

\begin{figure}
\includegraphics[width=8cm]{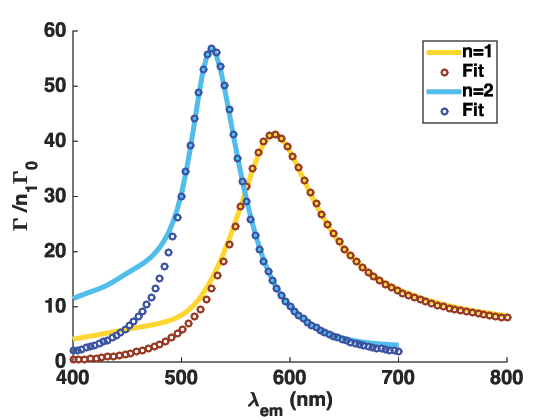}
\caption{Dipolar (n=1) and quadrupolar (n=2) contributions to the decay rate near a 80 nm gold MNP as a function of the emission wavelength. The emitter touches the MNP surface and is randomly oriented. Dots refer to  fits using Eq. (\ref{eq:PurcellLossDetuning}) with the following parameters. 
Dipolar  mode : $F_{p1}=41$, $\lambda_{1}=2\pi c/\omega_{1}=\SI{579}{\nano \meter}$, $Q_1=6$, $Im(V_1)/Re(V_1)=\SI{-0.07}{}$. Quadrupolar mode : $F_{p2}=34$, $\lambda_{2}=2\pi c/\omega_{2}=\SI{529}{\nano \meter}$, $Q_2=8.4$, $Im(V_2)/Re(V_2)=\SI{0.16}{}$.  the optical index of the surroundings medium is $n_1=1.5$.}
\label{fig:DecayLSP1Detuning}
\end{figure}

Moreover, detuning between the emission wavelength and the mode resonance $\omega_n$ leads to a Fano-like behaviour \cite{Sauvan-Lalanne:2013,Jiang-Perrin-Lalanne:2015}. Taking into account the (2n+1) LSPs degeneracy, the decay rate for a randomly oriented emitter obeys   
\begin{eqnarray}
\label{eq:PurcellLossDetuning}
\frac{\langle\Gamma_n\rangle}{n_1\Gamma_0} &=& \frac{(2n+1)}{3} F_p\frac{\left(\omega_n/\omega_{em}\right)^2}{1+ 4~Q^2\delta\widetilde{\omega}^2}
\left[1+2Q\delta \tilde{\omega} \frac{Im(V_n)}{Re(V_n)}\right] 
\;\nonumber\\
\delta\widetilde{\omega} &=&\frac{\omega_{em}-\omega_n}{\omega_{em}}
\end{eqnarray}
and generalizes expression (\ref{eq:DecayDetuning}) to lossy cavities. Figure \ref{fig:DecayLSP1Detuning} presents the dipolar and quadrupolar LSP contributions to the decay rate near a gold MNP. We observe an excellent agreement with the Fano-like  behaviour. From the fitting parameters, we extract the Purcell factor, quality factor and effective volume of the  dipolar mode:  $F_{p1}=41$, $Q_1=6$ and $V_1=\SI{6.4e-4}{\micro \meter ^3}=0.01\left(\lambda_{1}/n_1\right)^3$ respectively. Similarly, for the quadrupolar mode: $F_{p2}=34$, $Q_2=8.4$ and $V_2=\SI{8.2e-4}{\micro \meter ^3}=0.02\left(\lambda_{2}/n_1\right)^3$.
\subsubsection{Decay channels}
\begin{figure}
\includegraphics[width=8cm]{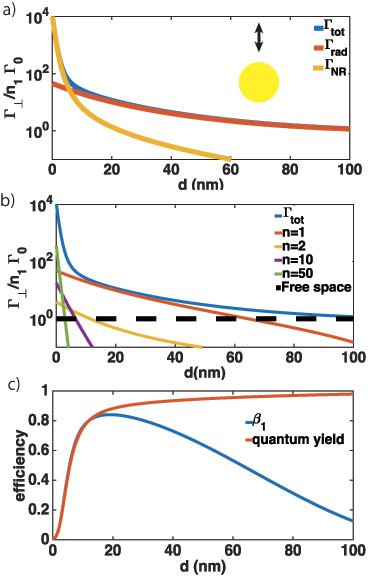}
\caption{Decay rate as a function of distance to a 80 nm gold particule. a) Total, radiative and non radiative decay rates. b) Contribution of the dipolar (n=1), quadrupolar (n=2) and high order modes (n=10 and 50) to the total decay rate. The horizontal dashed line is the free-space contribution. c) Coupling efficiency $\beta_1$ to the dipolar LSP and apparent quantum yield as a function of the distance to the particle. The emission wavelength is $\lambda_{em}=\SI{670}{\nano \meter}$ and the optical index of the surroundings medium is $n_1=1.5$. }
\label{fig:GamazLSP670}
\end{figure}
The Purcell factor quantifies the coupling efficiency into a given LSP mode but does not distinguish radiative from Joule losses, as for delocalized SPPs. 
Figure \ref{fig:GamazLSP670}a presents the radiative  and non radiative contributions to the decay rate near a gold MNP.  At very short distances, the decay rate is enhanced by several orders of magnitude. For distances below a few nanometers, the non local response of the metal permittivitty  (not included in this work) could lead to different values \cite{Leung:1990,Castanie-Boffety-Carminati:2010,Girard-Cuche:2015} but without significative change in the behaviour. In the very near-field, the main coupling mechanism is  
non-radiative energy transfer, responsible for fluorescence quenching \cite{anger06a}. Above $d=20$ nm, the total decay rate is still enhanced by a factor of about ten but is mainly radiative. In Fig. \ref{fig:GamazLSP670}b, we show the contribution of the dipolar, quadrupolar and high order LSP mode to the total decay rate. At short distances, high order modes play a dominant role since they are strongly confined near the particle surface, as revealed by their extremelly small effective volume (Fig. \ref{fig:Vnrj}) \cite{Mertens-Koenderink-Polman:2007,GCFIJMS:2009,OptExpGCF:2008,Kim:2015}. For distances of a few tens of nanometer, the radiative dipolar mode is responsible for the decay rate enhancement. This is summarized in Fig. \ref{fig:GamazLSP670}c where the coupling efficiency to the dipolar mode reaches $\beta_1=85\%$ at $d\approx20$ nm. The apparent quantum yield $\eta=\Gamma_{rad}/\Gamma_{tot}$ is governed by the dipolar mode in the near-field of the MNP as revealed by the two superimposed curve in Fig. \ref{fig:GamazLSP670}c.

\section{Conclusion}
In this work, we transposed the cQED Purcell factor to plasmonics. The Purcell factor characterizes the capability of a structure to modify and control the emission of a nearby emitter. Both optical cavity and plasmon modes present high Purcell factors that favor efficient excitation by a dipolar emitter. However it originates from high quality factors in case of optical cavities (but diffraction limited confinements) and on strongly subwavelength confinements for plasmonics (but low quality factors). Therefore the Purcell factor reveals the new paradigm opened by quantum plasmonics for achieving efficient light-matter interaction at the nanoscale. This permits a scale law approach profiting from the strong maturity of cQED concepts and adapt them to nanophotonics. We also discussed the presence of losses that are inevitable at the nanoscale. The Purcell factor includes Ohmic losses that are inherent to the excitation of plasmon in real metal so that it has to be manipulated with care. In the particular case of plasmonic waveguides, the coupling efficiency to a guided mode is not affected by propagation losses. Obviously, only low loss systems such that crystalline nanowires or nanoplatelets would permit realistics applications. Finally, hybrid plasmonic/nanophotonic configurations would profit from the concept of Purcell factor through a common description of the coupling mechanism. 

\subsection{Acknowledgments}
The research leading to these results has received funding from the  Agence Nationale de la Recherche
(grants QDOTICS ANR-12-BS-008 and PLACORE ANR-13-BS10-0007),  and the Conseil R\'egional de Bourgogne (PARI ACTION PHOTCOM).  
\appendix
\section{{Decay rate above a flat mirror}}
\subsection{Green's tensor and total decay rate}
\label{sect:G1D}
The Green's tensor above a mirror writes ${\bf G}={\bf G}_{0}+{\bf G}_{refl}$ where ${\bf G}_{0}$ is the Green's tensor of the infinite  homegeneous medium of optical index $n_1$ and ${\bf G}_{refl}$ describes the reflection on the mirror. It can be expressed thanks to Weyl expansion 
\begin{eqnarray}
\label{eq:Gslab}
&{\bf G}_{refl}({\bf r},{\bf r}_0)=\frac{i}{4\pi\varepsilon_1k_0^3}\int_0^\infty [g^s+g^p] e^{ik_{1z}(z+z_0)}dk_{\parallel} \\
&\text{with}\, k_1=n_1k_0\;, k_{1z}=(k_1^2-k^2_{\parallel})^{1/2} \;, [Im (k_{1z}) \ge 0] 
\nonumber
\end{eqnarray}
and, writing ${\bf r}-{\bf r}_0=(\rho,\varphi,z-z_0)$ in cylindrical coordinates ;  
\begin{eqnarray*}
&g^s_{xx}=k_1^2\frac{k_{\parallel}}{k_{1z}}r^s \left[\sin^2 \varphi J_0(k_{\parallel}\rho) +\cos (2\varphi)
\frac{J_1(k_{\parallel}\rho)}{k_{\parallel}\rho} \right] \\
&g^p_{xx}=-k_{1z}k_{\parallel}r^p\left[\cos^2 \varphi J_0(k_{\parallel}\rho) -\cos (2\varphi)
\frac{J_1(k_{\parallel}\rho)}{k_{\parallel}\rho} \right]\\
&g^s_{xy}=-k_1^2\frac{k_{\parallel}}{k_{1z}}r^s \sin \varphi \cos \varphi\left[ J_0(k_{\parallel}\rho) -2\frac{J_1(k_{\parallel}\rho)}{k_{\parallel}\rho}\right]\\
&g^p_{xy}=w1 k_{\parallel}r^p \sin \varphi \cos \varphi\left[ J_0(k_{\parallel}\rho) -2\frac{J_1(k_{\parallel}\rho)}{k_{\parallel}\rho}\right]\\
&g^s_{xz}=0 \;\;, g^p_{xz}=-ik^2_{\parallel}r^p \cos \varphi J_1(k_{\parallel}\rho) \\
&g^s_{yx}=g^s_{xy} \;\;, g^p_{yx}=g^p_{xy}\\
&g^s_{yy}=k_1^2\frac{k_{\parallel}}{k_{1z}}r^s \left[\cos^2 \varphi J_0(k_{\parallel}\rho) -\cos (2\varphi)
\frac{J_1(k_{\parallel}\rho)}{k_{\parallel}\rho} \right]\\
&g^p_{yy}=-k_{1z}k_{\parallel}r^p\left[\sin^2 \varphi J_0(k_{\parallel}\rho) +\cos (2\varphi)
\frac{J_1(k_{\parallel}\rho)}{k_{\parallel}\rho} \right]\\
&g^s_{yz}=0 \;\;, g^p_{yz}=-ik^2_{\parallel}r^p \sin \varphi J_1(k_{\parallel}\rho) \\
&g^s_{zx}=0 \;\;,
g^p_{zx}=ik^2_{\parallel}r^p \cos \varphi J_1(k_{\parallel}\rho) \\
&g^s_{zy}=0 \;\;,
g^p_{zy}=ik^2_{\parallel}r^p \sin \varphi J_1(k_{\parallel}\rho) \\
&g^s_{zz}=0 \;\;,
g^p_{zz}=\frac{k^3_{\parallel}}{k_{1z}}r^p J_0(k_{\parallel}\rho) 
\end{eqnarray*}
$r^s$ ($r^p$) refer to the fresnel reflexion coefficient on the slab for TE (TM) polarized light and $J$ are the cylindrical Bessel function.

The total decay rate of a dipolar emitter above follows Eq. \ref{eq:DecayGreen} and  writes above a slab 
\begin{eqnarray}
&\frac{\Gamma_u (d)}{n_1\Gamma_0}=\frac{3}{2n_1^{3/2}k_0^3}Re\int_0^\infty [g^s_{uu}+g^p] e^{2ik_{1z} d}dk_{\parallel}
\end{eqnarray}
with
\begin{eqnarray*}
&g^s_{xx}=g^s_{yy}=k_1^2\frac{k_{\parallel}}{2k_{1z}}r^s \\
&g^p_{xx}=g^p_{yy}=-\frac{1}{2} k_{1z}k_{\parallel}r^p \
&g^s_{zz}=0 \;\;,
g^p_{zz}=\frac{k^3_{\parallel}}{k_{1z}}r^p
\end{eqnarray*}
so that the total decay rate writes 
\begin{eqnarray}
\label{eq:DecaySpecA}
\frac{\Gamma_u(d)}{n_1\Gamma_0}&=&\int_0^\infty \mathcal{P}_{uu} (k_{\parallel})dk_{\parallel} \;,\text{with} \\
\mathcal{P}_{uu}(k_{\parallel})&=& \frac{3}{2n_1^3k_0^3}Re[(g^s_{uu}+g^p_{uu}) e^{2ik_{1z} d}]
\nonumber
\end{eqnarray}
that defines the dipolar emission power spectrum used in Eq. \ref{eq:GammaSpec}.
\subsection{SPP contribution near a lossless mirror}
\label{app:Lossless}
In this section, we consider a metal($\varepsilon_2$)/dielectric($\varepsilon_1$) single interface, and assume a lossless metal [$Im (\varepsilon_2)=0]$. The coefficient of reflexion writes 
\begin{eqnarray}
r^p=\frac{\varepsilon_2k_{1z}-\varepsilon_1k_{2z}}{\varepsilon_2k_{1z}+\varepsilon_1k_{2z}}
\end{eqnarray}
and presents a pole for $\varepsilon_2k_{1z}+\varepsilon_1k_{2z}=0$ leading to the (real) SPP wavector 
\begin{eqnarray}
k_{SPP}=k_1 \left(\frac{\varepsilon_2}{\varepsilon_1+\varepsilon_2}\right)^{1/2} \,.
\end{eqnarray}
The dipolar emission power spectrum near the SPP resonance can be assessed from an expansion of  $r_p$ near its pole
\begin{eqnarray}
r^p(k_{\parallel})&\underset{k_{SPP}}{\sim}&\frac{2\varepsilon_1\varepsilon_2}{\varepsilon_1^2-\varepsilon_2^2}\frac{k_{SPP}}{k_{\parallel}-k_{SPP}}
\end{eqnarray}
so that the SPP contribution to the decay rate is determined from the residu. 
We achieve  \cite{ford84}
\begin{eqnarray}
\nonumber
&\frac{\Gamma_\parallel (d)}{n_1\Gamma_0}=\frac{3\pi}{2\varepsilon_1} \frac{n^5_{SPP}}{(\varepsilon_1-\varepsilon_2)\vert\varepsilon_2\vert^{1/2}} e^{-2(\varepsilon_1/\vert\varepsilon_2\vert)^{1/2}k_{SPP}d} \\
\\
&\frac{\Gamma_\perp (d)}{n_1\Gamma_0}=\frac{3\pi}{n_1^3} \frac{n^5_{SPP}}{\varepsilon_1-\varepsilon_2}\vert\varepsilon_2\vert^{1/2} e^{-2(\varepsilon_1/\vert\varepsilon_2\vert)^{1/2}k_{SPP}d} 
\nonumber
\end{eqnarray}

\subsection{SPP contribution near a strongly lossy mirror}
\label{app:lossy}
\begin{figure}
\includegraphics[width=8cm]{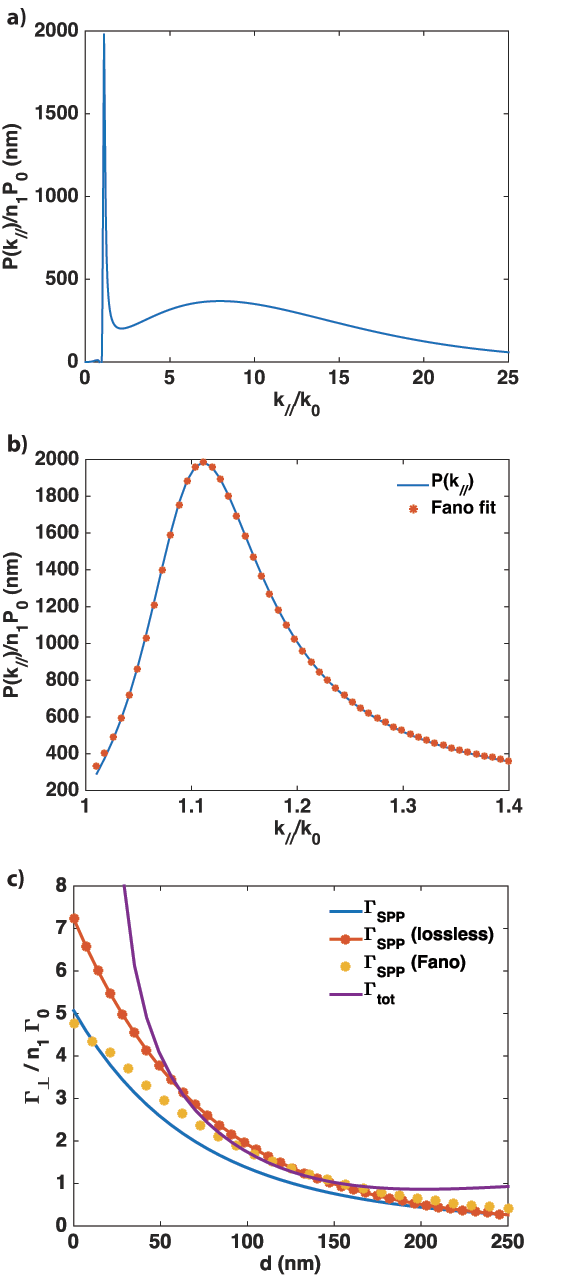}
\caption{a) Dipolar emission as a function of the wavector parallel to the surface $k_{\parallel}$ ($d=\SI{10}{\nano \meter}$). b) Resonant behaviour of the Au/air SPP. The Fano fit is peaked at the SPP effective index $n_{SPP}=1.10$ and with a FWHM $n_{SPP}^{''}=\SI{6.5e-2}{}$ ($L_{SPP}=\lambda_{em}/4\pi n_{SPP}^{''}=\SI{640}{\nano\meter}$). c) Estimation of the SPP contribution to the total decay rate (see text for details). The dipole is perpendicular to the surface. The emission wavelength is  $\lambda_{em}=\SI{525}{\nano \meter}$.}
\label{fig:SPPContrib525}
\end{figure}

We now consider the strongly lossy configuration at the emission wavelength $\lambda_{em}=\SI{525}{\nano \meter}$. 
Figure \ref{fig:SPPContrib525}a presents the dipolar emitter spectrum as a function of the in-plane wavector. Due to strong losses in the metal, lossy wave cannot be separated from the SPP resonance (compare to Fig. \ref{fig:PSlab}). Indeed,  the power spectrum presents a very broad resonance-like power spectrum at high $k_{\parallel}$. Therefore the dipolar emitter can couple to either the SPP or LSW, hence a Fano behaviour. In order to determine the SPP contribution to the total decay rate, we numerically integrate the emitted power in the range $1<k_{\parallel}/k_0<1.4$ (continuous line in Fig. \ref{fig:SPPContrib525}c). We also estimate the SPP behaviour assuming a Fano resonance as described in the next section (dots in Fig. \ref{fig:SPPContrib525}c). Finally, the  SPP contribution assuming a lossless metal and the total decay rate are also represented. We observe that the lossless model overestimates the SPP rate and could be even larger than the total decay rate (around $d\approx 100$ nm).  The SPP rate estimated from Fano profile or by direct numerical integration are in qualitative agreement but show some discrepencies, illustrating the diffculty to separate the SPP and LSW contributions.

\subsection{Scattering rate}
\label{app:Gscatt}
The scattering rate is achieved from the power scattering in the far-field zone. It can be estimated from asymptotic behaviour of the Green's tensor. 
We achieve 
\begin{eqnarray*}
\frac{\Gamma_{scatt}}{n_1\Gamma_0}=\int_0^{\pi/2}\sin\theta \sigma^{11}(\theta)d\theta+\frac{n_3}{n_1}\int_{\pi/2}^\pi \sin\theta  \sigma^{13}(\theta)d\theta 
\end{eqnarray*}
where the differential scattering cross-section expresses
\begin{eqnarray}
\label{eqsigma11x}
\sigma^{11}_{\parallel}(\theta)&=\frac{3}{4}-\frac{3}{8}\sin^2\theta+\frac{3}{8}\left(\vert r_p \vert ^2 \cos^2 \theta +\vert r_s \vert ^2\right) \\
\nonumber
&+\frac{3}{4} Re[(-r_p \cos \theta +r_s) e^{2ik_{1z}d} ] \\
\sigma^{11}_{\perp}(\theta)&=\frac{3}{4}+\frac{3}{4}\vert r_p \vert ^2 +\frac{3}{2} Re[r_p  e^{2ik_{1z}d} ]
\end{eqnarray}
above the film, for a dipole parallel or perpendicular to the surface, respectively and, 
\begin{eqnarray}
\sigma^{13}_{\parallel}(\theta)&=\frac{3}{8}\frac{\varepsilon_3}{\varepsilon_1}\vert e^{2ik_{1z}d} \vert \left(\vert t_p \vert ^2 +\vert \frac{k_1 t_s}{k_{1z}} \vert^2 \right) \cos^2 \theta \\
\sigma^{13}_{\perp}(\theta)&=\frac{3}{4}\frac{\varepsilon_3}{\varepsilon_1}\vert e^{2ik_{1z}d} \vert t_p \vert ^2 +\vert \frac{k_3 t_p}{k_{1z}} \vert^2 \cos^2 \theta \sin^2\theta
\end{eqnarray}
in the substrate. $t_p$ and $t_s$ refer to the Fresnel transmission coefficients for p and s polarized light, respectively.  This generalises the expression achieved for a thick mirror \cite{Chance:1975} to a finite metal slab or multilayer system. The difference betwen the radiative and scattering rates originates from the part of the radiative waves absorbed before achieving the far-field (see \emph{e.g.} Fig. \ref{fig:GamaAbs525}) and/or from SPCE leakage (see Fig. \ref{fig:Slab50Leak}).
 
\section{Fano profile}
\label{app:Fano}
Assuming a Fano profile of the emitted power $P(k_{\parallel})$ (see e.g. Fig. \ref{fig:GamaSlab}b), it follows \cite{Gallinet-Martin:2011}
\begin{eqnarray}
P(k_{\parallel})&=&\frac{P(k_{spp})}{q^2+b}\frac{(q\kappa_{spp}+k_{\parallel}-k_{spp})^2+b\kappa_{spp}^2}{(k_{\parallel}-k_{spp})^2+\kappa_{spp}^2}\\
&=&\frac{P(k_{spp})}{q^2+b}\frac{(x+q)^2+b}{1+x^2} \\
\text{with} &&x=\frac{k_{\parallel}-k_{spp}}{\kappa_{spp}}=\frac{k_{\parallel}/k_0-n_{spp}}{n''_{spp}}
\end{eqnarray}
where we note $\kappa_{spp}=1/2L_{SPP}$ the losses rate of the SPP. 

The SPP rate is then estimated from the integral of the emitted power. We first estimate the integral of the Fano resonance (the term $-1$ removes the background contribution) 
\begin{eqnarray}
I&=&\int_{-\infty}^{+\infty} \left[\frac{(x+q)^2+b}{1+x^2} -1\right] dx\\
\nonumber
&=&(q^2+b-1)\int_{-\infty}^{+\infty}  \frac{1}{1+x^2} dx+2q\int_{-\infty}^{+\infty}  \frac{x}{1+x^2} dx\\
&=&(q^2+b-1)\left[ atan(x)\right]_{-\infty}^{+\infty}+\left[qln(1+x^2)\right]_{-\infty}^{+\infty} \\
&=&(q^2+b-1)\pi 
\end{eqnarray}
and the SPP rate simplifies to 
\begin{eqnarray}
\Gamma_{SPP}&=&\int_{-\infty}^{+\infty} P(k_{\parallel})dk_{\parallel}\\
&=&\frac{1}{2L_{SPP}}\int_{-\infty}^{+\infty} P(x) dx\\
&=&\frac{P(k_{SPP})}{L_{SPP}}\frac{q^2+b-1}{q^2+b }\frac{\pi}{2}
\end{eqnarray}
so that we recover the expression (\ref{eq:SPPPurcell1}) for a Lorentzian profile ($q\rightarrow \infty$).

\section{Energy confinement of LSPs}
\label{sect:LSPVnrj}
The electrostatic potential associated to the $n^{th}$ LSP mode of a spherical MNP follows \cite{Jackson:1998} 
\begin{eqnarray}
\hspace{-0.35cm}
\Phi_{n,m}=\sum\limits_{m=-n}^n C_{n,m}\left(\frac{r}{R}\right)^nP_n^m(\cos\theta)e^{im\varphi} \,,r<R\\
\hspace{-0.35cm}
\Phi_{n,m}=\sum\limits_{m=-n}^n C_{n,m}\left(\frac{R}{r}\right)^{n+1}P_n^m(\cos\theta)e^{im\varphi} \,, r \ge R
\end{eqnarray}
with the normalisation constant 
\begin{eqnarray}
 C_{n,m}=(-1)^m\sqrt{\frac{2n+1}{4\pi}\frac{(n-m)!}{(n+m)!}}
\end{eqnarray}
and $P_n^m$ is the associated Legendre polynomial.
The boundary conditions imposes 
\begin{eqnarray}
n\varepsilon_m+(n+1)\varepsilon_1=0
\end{eqnarray}
 that fixes the resonances of the $n^{th}$ LSP. If the dielectric constant of the metal $\varepsilon_m$ follows a Drude behaviour, the resonance angular frequency obeys 
$\omega_n=\omega_p\sqrt{n/n+(n+1)\varepsilon_1}$. 
The $n^{th}$ LSP presents a degeneracy $g_n=(2n+1)$ and the electric field of the (n,m) mode writes 
\begin{eqnarray}
{\bf E}_{n,m}=C_{n,m}\left(\frac{r}{R}\right)^{n-1} 
\left[
\begin{array}{l}
-nP_n^m(\cos\theta)e^{im\varphi}\;{\bf e}_r\\
\frac{1}{\sin \theta}\left[(n+1)\cos\theta P_{n}^m(\cos\theta)-(n-m+1)P_{n+1}^m\right]e^{im\varphi} \;{\bf e}_\theta\ \;,( r< R)\\
\frac{im}{\sin \theta}P_n^m(\cos\theta)e^{im\varphi} \;{\bf e}_\varphi\
\end{array}
\right.
\\ \nonumber \\
{\bf E}_{n,m}=C_{n,m}\left(\frac{R}{r}\right)^{n+2} 
\left[
\begin{array}{l}
-(n+1)P_n^m(\cos\theta)e^{im\varphi}\;{\bf e}_r\\
\frac{1}{\sin \theta}\left[(n+1)\cos\theta P_{n}^m(\cos\theta)-(n-m+1)P_{n+1}^m\right]e^{im\varphi} \;{\bf e}_\theta\ \;, (r\ge R)\\
-\frac{im}{\sin \theta}P_n^m(\cos\theta)e^{im\varphi} \;{\bf e}_\varphi\
\end{array}
\right.
\end{eqnarray}

The effective volume is expressed extrapolating the cQED description to a dispersive medium
\begin{eqnarray}
V_{n,m}&=&\frac{\int  U_{n,m}({\bf r}) d{\bf r}}{max[\varepsilon_0\varepsilon_1~|{\bf E_{n,m=0}}|^2]} \;, \\
U_{n,m}({\bf r})&=&\frac{\partial[\omega\varepsilon_0\varepsilon({\bf r},\omega)]}{\partial \omega} ~|{\bf E_{n}}({\bf r})|^2
\nonumber
\label{eq:Vlm}
\end{eqnarray} 
Note that,since the mode is degenerated, we normalized the field with respect to maximum of the field considering \emph{all} the degenerated modes ($Max|{\bf E_{n}}({\bf r})|^2$ and not ($Max|{\bf E_{n,m}}({\bf r})|^2$  in the denominator of Eq. \ref{eq:Vlm}). Assuming a lossless Drude metal, it comes
\begin{eqnarray}
\frac{\partial[\omega\varepsilon_m(\omega)]}{\partial \omega} \bigg\vert_{\omega=\omega_n}=2+\frac{n+1}{n}\varepsilon_1
\end{eqnarray}
so that the effective volume is for m=0  \cite{Khurgin-Sun:2009}
\begin{eqnarray}
V_{n,0} =\frac{6V_0}{(2n+1)(n+1)}\left[1+\frac{n}{(n+1)\varepsilon_1} \right]
\end{eqnarray}
For $m \ne 0$, the integration over $\theta$ is numerically evaluated (see Fig. \ref{fig:Vlm}). 
We note a discrepancy between the mode volume deduced from Purcell factor ($V_{Purcell}$) and the mode volume estimated from energy mode confinement assuming a lossless metal ( $V_{n,m}$). Indeed, we have \cite{GCFIJMS:2009}
\begin{eqnarray}
\nonumber
V_{n}^{Purcell} =\frac{3V_0}{(2n+1)}\left[1+\frac{n}{(n+1)\varepsilon_1} \right]
\end{eqnarray}
so that the ratio 
\begin{eqnarray}
\frac{V_{Purcell}}{V_{energy}}=\frac{n+1}{2}
\end{eqnarray}
depends on the mode number $n$ only.


\begin{thebibliography}{100}
\expandafter\ifx\csname url\endcsname\relax
  \def\url#1{{\tt #1}}\fi
\expandafter\ifx\csname urlprefix\endcsname\relax\def\urlprefix{URL }\fi
\providecommand{\eprint}[2][]{\url{#2}}

\bibitem{Hell:2007}
Hell S~W 2007 {\em Science\/} {\bf 316} 1153

\bibitem{Celebrano-Sandoghdar:2011}
Celebrano M, Kukura P, Renn A and Sandoghdar V 2011 {\em Nature Photonics\/}
  {\bf 5} 95--98

\bibitem{Tamarat2000}
Tamarat P, Maali A, Lounis B and Orrit M 2000 {\em The Journal of Physical
  Chemistry A\/} {\bf 104} 1--16

\bibitem{Vahala:2003}
Vahala K~J 2003 {\em Nature\/} {\bf 424} 839--846

\bibitem{Betzig1993}
Betzig E and Chichester R~J 1993 {\em Science\/} {\bf 262} 1422--1425

\bibitem{Veerman-Garcia-Kuipers-VanHulst:1999}
Veerman J~A, Garcia-Parajo M~F, Kuipers L and {van Hulst} N~F 1999 {\em J. of
  Microscopy\/} {\bf 194} 477--482

\bibitem{Sick2001}
Sick B, Hecht B, Wild U~P and Novotny L 2001 {\em Journal of Microscopy\/} {\bf
  202} 365--373

\bibitem{OptExpMolendaGCF:2005}
Molenda D, {Colas des Francs} G, Fischer U, Rau N and Naber A 2005
  {\em Optics Express\/} {\bf 13} 10688

\bibitem{Wenger-Rigneault:2010}
Wenger J and Rigneault H 2010 {\em International Journal of Molecular
  Science\/} {\bf 11} 206--221

\bibitem{Brun-Drezet-Mariette-Chevalier-Woelh-Huant:2003}
Brun M, Drezet A, Mariette H, Chevalier N, Woehl J~C and Huant S 2003 {\em
  Europhys. Lett.\/} {\bf 64} 634--640

\bibitem{bharadwaj07a}
Bharadwaj P, Anger P and Novotny L 2007  {\bf 18} 044017

\bibitem{Grynberg-Aspect-Fabre:2010}
Grynberg G, Aspect A and Fabre C 2010 {\em Quantum optics\/} (Paris: Cambridge
  University Press)

\bibitem{TanjiSuzuki-Vuletic:2011}
Tanji-Suzuki H, Leroux I, Schleier-Smith M, Cetina M, Grier A, Simon J and
  Vuletic V 2011 {\em Advances in Atomic, Molecular, and Optical Physics\/}
  {\bf 60} 201--240

\bibitem{Purcell:1946}
Purcell E 1946 {\em Physical Review\/} {\bf 69} 681

\bibitem{Agio:2012}
Agio M 2012 {\em Nanoscale\/} {\bf 4} 692--706

\bibitem{Sandoghdar-Agio:2012}
Sandoghdar V, Agio M, Chen X, Gotzinger S and Lee K 2012 {\em Antennas, quantum
  optics and near-field microscopy\/} (Cambrid)

\bibitem{Lodahl:2015}
Lodahl P, Mahmoodian S and Stobbe S 2015 {\em Review of Modern Physics\/} {\bf
  87} 347--400

\bibitem{vanExter-Woerdman:1996}
van Exter M~P, Nienhuis G and Woerdman J~P 96 {\em Physical Review A\/} {\bf
  54} 3553--3558

\bibitem{Brinks-vanHulst:2013}
Brinks D, Castro-Lopeza M, Hildner R and van Hulst N 2013 {\em PNAS\/}

\bibitem{Lalanne-Beveratos:2013}
Elvira D, Braive R, Beaudoin G, Sagnes I, Hugonin J, Abram I, Robert-Philip,
  Lalanne P and Beveratos A 2013 {\em Applied Physics Letters\/} {\bf 103}
  061113

\bibitem{Chang-Sorensen-Hemmer-Lukin:2006}
Chang D, S{\"o}rensen A, Hemmer P and Lukin M 2006 {\em Physical Review
  Letters\/} {\bf 97} 053002

\bibitem{Waks-Sridharan:2010}
Waks E and Sridharan D 2010 {\em Physical Review A\/} {\bf 82} 043845

\bibitem{Buckley-Vukovic:2012}
Buckley S, Rivoire K and Vuckovic J 2012 {\em Report on Progress in Physics\/}
  {\bf 75} 126503

\bibitem{Hummer-Garciavidal:2013}
H{\"u}mmer T, Garc{\'i}a-Vidal F~J, Mart{\'i}n-Moreno L and Zueco D 2013 {\em
  Physical Review B\/} {\bf 87} 115419

\bibitem{Tame-Maier:2013}
Tame M~S, McEnery K~R, Ozdemir S~K, Lee J, Maier S~A and Kim M~S 2013 {\em
  Nature Physics\/} {\bf 9} 329--340

\bibitem{Hakami-Wang-Zubairy:2014}
Hakami J, Wang L and Zubairy M 2014 {\em Physical Review A\/} {\bf 89} 053835

\bibitem{Torma-Barnes:2015}
T{\"o}rm{\"a} P and Barnes W 2015 {\em Rep. Prog. Phys.\/} {\bf 78} 013901

\bibitem{Rousseaux-GCF-Guerin:2015}
Rousseaux B, Dzsotjan D, {Colas des Francs} G, Jauslin H, Couteau C
  and Gu\'erin S 2015 {\em submitted\/}

\bibitem{Gerard:2003}
G\'erard J~M 2003 {\em Topics Appl. Phys.\/} {\bf 90} 269--315

\bibitem{Yalla-Hakuta:2012}
Yalla R, Kien F~L, Morinaga M and Hakuta K 2012 {\em Physical Review Letters\/}
  {\bf 109} 063602

\bibitem{Claudon-Gregersen-Lalanne-Gerard:13}
Claudon J, Gregersen N, Lalanne P and Gerard J~M 2013 {\em ChemPhysChem\/} {\bf
  14} 2393--2402

\bibitem{JunPhD:2010}
Jun Y 2010 {\em Plasmonic control of light emission: tailoring light emission
  properties with metal nanostructures\/} Ph.D. thesis Stanford University

\bibitem{Woldeyohannes-John:2003}
Woldeyohannes M and John S 2003 {\em J. Opt. B: Quantum Semiclass. Opt.\/} {\bf
  5} R43--R82

\bibitem{Chen-Brandes:2009}
Chen Y~N, Chen G~Y, Chuu D~S and Brandes T 2009 {\em Physical Review A\/} {\bf
  79} 033815

\bibitem{Nomura-Arakawa:2008}
Nomura M, Iwamoto S, Kumagai N and Arakawa Y 2008 {\em Physica E\/} {\bf 40}
  1800--1803

\bibitem{metiu84}
Metiu H 1984 Surface enhanced spectroscopy {\em Progress in Surface Science\/}
  vol~17 ed Prigogine I and Rice S~A (New York: Pergamon Press) pp 153--320

\bibitem{Girard1995l}
Girard C, Martin O~J~F and Dereux A 1995 {\em Physical Review Letters\/} {\bf
  75} 3098--3111

\bibitem{JCPGCF:2002}
{Colas des Francs} G, Girard C and Dereux A 2002 {\em Journal of
  Chemical Physics\/} {\bf 117} 4659--4666

\bibitem{Knoll-Scheel-Welsch:01}
Knoll L, Scheel S and Welsch D 2001 {\em Coherence and Statistics of Photons
  and Atoms\/} (John Wiley \& Sons, Inc.) chap QED in dispersing and absorbing
  media

\bibitem{Vries1998}
de~Vries P, van Coeverden D~V and Lagendijk A 1998 {\em Review of Modern
  Physics\/} {\bf 70} 447--466

\bibitem{moskovits05}
Moskovits M 2005 {\em Journal of Raman Spectroscopy\/} {\bf 36} 485--496

\bibitem{LeRu-Etchegoin:12}
Ru E~L and Etchegoin P 2012 {\em Annu. Rev. Phys. Chem.\/} {\bf 63} 65--87

\bibitem{LeRu-Etchegoin:08}
Ru E~L and Etchegoin P 2008 {\em Principles of surface-enhanced Raman
  spectroscopy and related plasmonic effects\/} (Elsevier Science)

\bibitem{Maier:2006}
Maier S 2006 {\em Optics Express\/} {\bf 14} 1957--1964

\bibitem{Fort-Gresillon:2005}
Fort E and Gr\'esillon S 2008 {\em Journal of Physics D: Applied Physics\/}
  {\bf 41} 013001

\bibitem{Lakowicz2005}
Lakowicz J~R 2005 {\em Analytical Biochemistry\/} {\bf 337} 171--194

\bibitem{Kuhn-Sandoghdar:2006}
K{\"u}hn S, Hakanson U, Rogobete L and Sandoghdar V 2006 {\em Phys. Rev.
  Lett.\/} {\bf 97} 017402

\bibitem{Anger-Bharadwaj-Novotny:2006}
Anger P, Bharadwaj P and Novotny L 2006 {\em Phys. Rev. Lett.\/} {\bf 96}
  113002

\bibitem{Viste-Plain:10}
Viste P, Plain J, Jaffiol R, Vial A, Adam P~M and Royer P 2010 {\em ACS Nano\/}
  {\bf 4} 759--764

\bibitem{Derom-GCF:2013}
Derom S, Berthelot A, Pillonnet A, Benamara O, Jurdyc A, Girard C and
  {Colas des Francs} G 2013 {\em Nanotechnology\/} {\bf 24} 495704

\bibitem{loo04}
Loo C, Lin A, Hirsch L, Lee M~H, Barton J, Halas N, West J and Drezek R 2004
  {\em Technol. Cancer. Res. T.\/} {\bf 3} 33--40

\bibitem{Wang-Chuang-AnnieHo:12}
Wang L~S, Chuang M~C and Ho J~A~A 2012 {\em International Journal of
  Nanomedicine\/} {\bf 7} 4679--4695

\bibitem{Brokman-Hermier:2004}
Brokmann X, Coolen L, Dahan M and Hermier J~P 2004 {\em Physical Review
  Letters\/} {\bf 93} 107403

\bibitem{Buchler-Kalkbrenner-Hettich-Sandoghdar:2005}
Buchler B~C, Kalkbrenner T, Hettich C and Sandoghdar V 2005 {\em Physical
  Review Letters\/} {\bf 95} 63004

\bibitem{Mertens-Koenderink-Polman:2007}
Mertens H, Koenderink A and Polman A 2007 {\em Physical Review B\/} {\bf 76}
  115123

\bibitem{Carminati-deWilde:2015}
Carminati R, Caze A, Cao D, Peragut F, Krachmalnicoff V, Pierrat R and Wilde
  Y~D 2015 {\em Surface Science Reports\/} {\bf 70} 1--41

\bibitem{CPLGirardGCF:2005}
Girard C, Martin O, L\'ev\^eque G, {Colas des Francs} G and Dereux A
  2005 {\em Chemical Physics Letters\/} {\bf 404} 44

\bibitem{stefani07}
Stefani F~D, Vasilev K, Bocchio N, Gaul F, Pomozzi A and Kreiter M 2007 {\em
  New J. Phys.\/} {\bf 9} 21

\bibitem{fu07}
Fu Y, Zhang J and Lakowicz J~R 2007  {\bf 447} 96--100

\bibitem{Greffet-Dubertret:2015}
Ji B, Giovanelli E, Habert B, Spinicelli P, Nasilowski M, Xu X, Lequeux N,
  Hugonin J~P, Marquier F, Greffet J and Dubertret B 2015 {\em Nature
  Nanotechnology\/} {\bf 10} 170--175

\bibitem{Cang-Xhang:2013}
Cang H, YLiu, Wang Y, Yin X and Zhang X 13 {\em Nano Lett.\/} {\bf 2013}
  5949--5953

\bibitem{Karaveli-Zia:2011}
Karaveli S and Zia R 2011 {\em Physical Review Letters\/} {\bf 106} 193004

\bibitem{Taminiau-Zia:2012}
Taminiau T, Karaveli S, van Hulst N and Zia R 2012 {\em Nature
  Communications\/} {\bf 3} 979

\bibitem{HechtScience:2005}
M{\"u}hlschlegel P, Eisler H~J, Martin O~J~F, Hecht B and Pohl D~W 2005 {\em
  Science\/} {\bf 308} 1607 -- 1609

\bibitem{Taminiau-Segerink-vanHulst:2007}
Taminiau T~H, Segerink F~B and van Hulst N~F 2007 {\em IEEE Transactions on
  antennas and propagation\/} {\bf 55} 3010--3017

\bibitem{PRBHuang:2008}
Huang C, Bouhelier A, {Colas des Francs} G, Bruyant A, Gu\'enot A,
  Finot E, Weeber J~C and Dereux A 2008 {\em Physical Review B\/} {\bf 78}
  155407

\bibitem{Bonod2010}
Bonod N, Devilez A, Rolly B, Bidault S and Stout B 2010 {\em Physical Review
  B\/} {\bf 82} 115429

\bibitem{Bharadwaj-Deutsch-Novotny:2009}
Bharadwaj P, Deutsch B and Novotny L 2009 {\em Adv. Opt. Phot.\/} {\bf 1}
  438--483

\bibitem{Greffet-Laroche-Marquier:2010}
Greffet J~J, Laroche M and Marquier F 2010 {\em Physical Review Letters\/} {\bf
  105} 117701

\bibitem{Krasnok-Belov:2015}
Krasnok A, Slobozhanyuk A, Simovski C, Tretyakov S, Poddubny A, Miroshnichenko
  A, Kivshar Y and Belov P 2015 {\em Scientific Reports\/} {\bf 5} 12956

\bibitem{Delga-GarciaVidal:2014}
Delga A, Feist J, Bravo-Abad J and Garcia-Vidal F~J 2014 {\em Physical Review
  Letters\/} {\bf 112} 253601

\bibitem{Akselrord-Mikkelsen:2014}
Akselrod G, Argyropoulos C, Hoang T, Ciraci C, CFang, Huang J, Smith D and
  Mikkelsen M 2014 {\em Nature Photonics\/} {\bf 8} 835--840

\bibitem{Koenderink:12}
Arango F~B, Kwadrin A and Koenderink A 2012 {\em ACS Nano\/} {\bf 6}
  10156--10167

\bibitem{Schietinger-Barth-Aichele-Benson:2009}
Schietinger S, Barth M, Aichele T and Benson O 2009 {\em Nano Letters\/} {\bf
  9} 1694--1698

\bibitem{Marty-Arbouet-Paillard-Girard-GCF:2010}
Marty R, Arbouet A, Paillard V, Girard C and {Colas des Francs} G
  2010 {\em Physical Review B\/} {\bf 82} 081403(R)

\bibitem{MallekZouari:2010}
Mallek-Zouari I, Buil S, Quelin X, Mahler B, Dubertret B and Hermier J~P 2010
  {\em Applied Physics Letters\/} {\bf 97} 053109

\bibitem{Celebrenao-Sandoghdar:2010}
Celebrano M, Lettow R, Kukura P, Agio M, Renn A, Gotzinger S and Sandoghdar V
  2010 {\em Optics Expres\/} {\bf 18} 13829--13835

\bibitem{Cuche2010}
Cuche A, Mollet O, Drezet A and Huant S 2010 {\em Nano Letters\/} {\bf 10}
  4566--4570

\bibitem{Busson-Bonod-Bidault:2012}
Busson M~P, Rolly B, Stout B, Bonod N and Bidault S 2012 {\em Nature
  Communications\/} {\bf 3} 962

\bibitem{Mollet2012}
Mollet O, Huant S, Dantelle G, Gacoin T and Drezet A 2012 {\em Phys. Rev. B\/}
  {\bf 86} 045401

\bibitem{GU-MARTIN:2012}
Gu Y, Wang L, Ren P, Zhang J, Zhang T, Martin O~J~F and Gong Q 2012 {\em
  NanoL\/} {\bf 12} 2488--2493

\bibitem{Rockstuhl:2014}
Filter R, Slowik K, Straubel J, Lederer F and Rockstuhl C 2014 {\em Optics
  Letters\/} {\bf 39} 1246--1249

\bibitem{Nerkararyan-Bozhevolnyi:2014}
Nerkararyan K and Bozhevolnyi S 2014 {\em Optics Letters\/} {\bf 39} 1617--1620

\bibitem{Beveratos:2014}
Beveratos A, Gerard I~A~A and Robert-Philip I 2014 {\em The European Physical
  Journal D\/} {\bf 68} 377

\bibitem{Grange-Auffeves:2015}
Grange T, Hornecker G, Hunger D, Poizat J, Gerard J, Senellart P and Auffeves A
  2015 {\em Physical Review Letters\/} {\bf 114} 193601

\bibitem{Bergman2003}
Bergman D~J and Stockman M~I 2003 {\em Physical Review Letters\/} {\bf 90}
  027402

\bibitem{Protsenko:2005}
Protsenko I~E, Uskov A~V, Zaimidoroga O~A, Samoilov V~N and {O'Reilly} E 2005
  {\em Physical Review A\/} {\bf 71} 063812

\bibitem{Berini-deLeon:2012}
Berini P and Leon I~D 2012 {\em Nature Photonics\/} {\bf 6} 16--24

\bibitem{Winter-Wedge-Barnes:2006}
Winter G, Wedge S and Barnes W~L 2006 {\em New Journal of Physics\/} {\bf 8}
  125

\bibitem{deLeon-Berini:2008}
{De Leon} I and Berini P 2008 {\em Physical Review B\/} {\bf 78} 161401

\bibitem{ColasdesFrancsOptExp:2010}
{Colas des Francs} G, Bramant P, Grandidier J, Bouhelier A, Weeber
  J~C and Dereux A 2010 {\em Optics Express\/} {\bf 18} 16327--16334

\bibitem{GrandidierJMicrosc:2010}
Grandidier J, {Colas des Francs} G, Massenot S, Bouhelier A, Weeber
  J~C, Markey L and Dereux A 2010 {\em Journal of Microscopy\/} {\bf 239}
  167--172

\bibitem{Andrew-Barnes:2000}
Andrew P and Barnes W~L 2000 {\em Science\/} {\bf 290} 785--788

\bibitem{MartinCano-MartinMoreno-GarciaVidal-Moreno:2010}
{Martin-Cano} D, {Martin-Moreno} L, {Garcia-Vidal} F and Moreno E 2010 {\em
  Nano Letters\/} {\bf 10} 3129--3134

\bibitem{ZuritaSanchez:2013}
{Gonzaga-Galeana} J and {Zurita-Sanchez} J 2013 {\em The journal of physical
  chemistry\/} {\bf 139} 244302

\bibitem{Karanikolas-Bradley:2014}
Karanikolas V, Marocico C and Bradley A~L 2014 {\em Physical Review A\/} {\bf
  89} 063817

\bibitem{Krachmalnicoff:2015}
Bouchet D, Cao D, Carminati R, Wilde Y~D and Krachmalnicoff V 2015 {\em
  arxiv\/} {\bf 1507.04235}

\bibitem{Fleischhauer:2010}
Dzsotjan D, Sorensen A~S and Fleischhauer M 2010 {\em Physical Review B\/} {\bf
  82} 075427

\bibitem{GonzalezTuleda-GarciaVidal:2011}
Gonzalez-Tudela A, Martin-Cano D, Moreno E, Martin-Moreno L, Tejedor C and
  Garcia-Vidal F~J 2011 {\em Physical Review Letters\/} {\bf 106} 020501

\bibitem{ford84}
Ford G~W and Weber W~H 1984 {\em Physics Reports\/} {\bf 113} 195--287

\bibitem{Barnes1998b}
Barnes W 1998 {\em Journal of Modern Optics\/} {\bf 45} 661--699

\bibitem{Johnson-Christy:1972}
Johnson P and Christy R 1972 {\em Physical Review B\/} {\bf 6} 4370--4379

\bibitem{Chance:1975}
Chance R~R, Prock A and Silbey R 1975  {\bf 62} 2245--2253

\bibitem{Hinds:1991}
Hinds E 1991 {\em Advances in atomic, molecular, and optical physics\/} {\bf
  28} 237--289

\bibitem{Barthes-GCF-Bouhelier-Weeber-Dereux:2011}
Barthes J, {Colas des Francs} G, Bouhelier A, Weeber J~C and Dereux A
  2011 {\em Physical Review B (Brief Reports)\/} {\bf 84} 073403

\bibitem{Jung:2012}
Siahpoush V, Sondergaard T and Jung J 2012 {\em Physical Review B\/} {\bf 85}
  075035

\bibitem{ColasdesFrancsPRB:2009}
{Colas des Francs} G, Grandidier J, Massenot S, Bouhelier A, Weeber
  J~C and Dereux A 2009 {\em Physical Review B\/} {\bf 80} 115419

\bibitem{Aouani-Wenger:2011}
Aouani H, Mahboub O, Bonod N, Devaux E, Popov E, Rigneault H, Ebbesen T and
  Wenger J 2011 {\em Nano Letters\/} {\bf 11} 637--644

\bibitem{Choy-Loncar:2013}
Choy J~T, Bulu I, Hausmann B~M, Janitz E, Huang I~C and Lonc{\~a}r M 2013 {\em
  Applied Physics Letters\/} {\bf 103} 161101

\bibitem{Kumar-Dubertret-GCF:2015}
Kumar A, Weeber J~C, Bouhelier A, Eloi F, Buil S, Quelin X, Nasilowski M,
  Dubertret B, Hermier J and  {Colas des Francs} G 2015 {\em
  Scientific Reports\/} {\bf 5} 16796

\bibitem{Anemogiannis-Glytsis-Gaylord:1999}
Anemogiannis E, Glytsis E~N and Gaylord T~K 1999 {\em Journal of Lightwave
  Technology\/} {\bf 17} 929--941

\bibitem{Mollet-Drezet:2012}
Mollet O, Huant S, Dantelle G, Gacoin T and Drezet A 2012 {\em Phys. Rev. B\/}
  {\bf 86} 045401

\bibitem{Miroshnichenko:2010}
Miroshnichenko A~E, Flach S and Kivshar Y~S 2010 {\em Review of Modern
  Physics\/} {\bf 82} 2257--2298

\bibitem{Gallinet-Martin:2011}
Gallinet B and Martin O 2011 {\em Physical Review B\/} {\bf 83} 235427

\bibitem{Drezet2008}
Drezet A, Hohenau A, Koller D, Stepanov A, Ditlbacher H, Steinberger B,
  Aussenegg F, Leitner A and Krenn J 2008 {\em Materials Science and
  Engineering B\/} {\bf 149} 220--229

\bibitem{Bouhelier-ColasdesFrancs-Grandidier:2012}
Bouhelier A, {Colas des Francs} G and Grandidier J 2012 {\em
  Plasmonics, From Basics to Advanced Topics\/} ({\em Springer Series in
  Optical Sciences\/} vol 167) (Springer) chap Surface plasmon imaging, pp
  225--268

\bibitem{Lakowicz5:2005}
Lakowicz J~R 2005 {\em Analytical Biochemistry\/} {\bf 337} 171--194

\bibitem{Raether1986}
Raether H 1986 {\em Surface Plasmons on Smooth and Rough Surfaces and on
  Gratings\/} (Springer Verlag)

\bibitem{NanoLettWeeberGCF:2007}
Weeber J~C, Bouhelier A, {Colas des Francs} G, Markey L and Dereux A
  2007 {\em Nanoletters\/} {\bf 7} 1352

\bibitem{Gonga-Vukovic:2007}
Gonga Y and Vuckovic J 2007 {\em Applied Physics Letters\/} {\bf 90} 033113

\bibitem{Derom-Hermier-GCF:2014}
Derom S, Bouhelier A, Kumar A, Leray A, Weeber J~C, Buil S, Quelin X, Hermier J
  and  {G {Colas des Francs}} 2014 {\em Physical Review B\/} {\bf 89}
  035401

\bibitem{Barthes-Bouhelier-Dereux-GCF:2013}
Barthes J, Bouhelier A, Dereux A and {Colas des Francs} G 2013 {\em
  Scientific Reports\/} {\bf 3} 2734

\bibitem{Takahara:1997}
Takahara J, Yamagishi S, Taki H, Morimoto A and Kobayashi T 1997 {\em Optics
  Letters 22\/} {\bf 22} 475--477

\bibitem{Chang-Sorensen-Hemmer-Lukin:2007}
Chang D, S{\"o}rensen A, Hemmer P and Lukin M 2007 {\em Physical Review B\/}
  {\bf 76} 35420

\bibitem{Kolesov2009}
Kolesov R, Grotz B, Balasubramanian G, St{\"o}hr R~J, Nicolet A~A~L, Hemmer
  P~R, Jelezko F and Wrachtrup J 2009 {\em Nature Physics\/} {\bf 5} 470--474

\bibitem{Huck-Andersen:2009}
Huck A, Kumar S, Shakoor A and Andersen U~L 2009 {\em Physical Review
  Letters\/} {\bf 106} 096801

\bibitem{Gruber-Krenn:2012}
Gruber C, Kusar P, Hohenau A and Krenn J~R 2012 {\em Applied Physics Letters\/}
  {\bf 100} 221102

\bibitem{Ropp-Waks:2013}
Ropp C, Cummins Z, Nah S, Fourkas J~T, Shapiro B and Waks E 2013 {\em Nature
  Communications\/} {\bf 4} 1447

\bibitem{Bozhelvonyi-Quidant:2015}
Bermudez-Urena E, Gonzalez-Ballestero C, Geiselmann M, Marty R, Radko I,
  Holmgaard T~A, Alaverdyan Y, Moreno E, Garcia-Vidal F, Bozhevolnyi S and
  Quidant R 2015 {\em Nature Communications\/} {\bf 6} 7883

\bibitem{Norris:2015}
Kress S, Antolinez F, Richner P, Jayanti S, Kim D, Prins F, Riedinger A,
  Fischer M, Meyer S, McPeak K, Poulikakos D and Norris D 2015 {\em
  NanoLetters\/}  ASAP

\bibitem{Wei-Xu:2012}
Wei H and Xu H 2012 {\em Nanophotonics\/} {\bf 1} 155--169

\bibitem{Ditlbacher2005}
Ditlbacher H, Hohenau A, Wagner D, Kreibig U, Rogers M, Hofer F, Aussenegg F~R
  and Krenn J~R 2005 {\em Physical Review Letters\/} {\bf 95} 257403

\bibitem{Laroche-Vial-Roussey:2007}
Laroche T, Vial A and Roussey M 2007 {\em Applied Physics Letters\/} {\bf 91}
  123101

\bibitem{Song-Dujardin-Zhang-GCF:2011}
Song M, Bramant P, Bouhelier A, Sharma J, Dujardin E, Zhang D and
  {Colas des Francs} G 2011 {\em ACS Nano\/} {\bf 5} 5874--5880

\bibitem{Viarbitskaya:2013}
Viarbitskaya S, Teulle A, Cuche A, Sharma J, Girard C, Dujardin E and Arbouet A
  2013 {\em Applied Physics Letters\/} {\bf 103} 131112

\bibitem{Viarbitskaya-Dujardin:2013}
Viarbitskaya S, Teulle A, Marty R, Sharma J, Girard C, Arbouet A~A and Dujardin
  E 2013 {\em Nature Materials\/} {\bf 12} 426--432

\bibitem{Chang-Sorensen-Demler-Lukin:2007}
Chang D, S{\"o}rensen A, Demler E and Lukin M 2007 {\em Nature Physics\/} {\bf
  3} 807--812

\bibitem{Chen-Nori:2011}
Chen G~Y, Lamber N, Chou C~H, Chen Y~N and Nori F 2011 {\em Physical Review
  B\/} {\bf 84} 045310

\bibitem{GrandidierNanoLet:2009}
Grandidier J, {Colas des Francs} G, Massenot S, Bouhelier A, Markey
  L, Weeber J, Finot C and Dereux A 2009 {\em Nano Letters\/} {\bf 9}
  2935--2939

\bibitem{KenaCohen-Maier:2013}
K\'ena-Cohen S, Stavrinou P, Bradley D and Maier S 2013 {\em Nano Letters\/}
  {\bf 13} 1323--1329

\bibitem{Paul-Nordlander-Link:2014}
Paul A, Zhen Y~R, Wang Y, Chang W~S, Xia Y, Nordlander P and Link S 2014 {\em
  NanoLetters\/} {\bf 14} 3628--3633

\bibitem{Snyder-Love:1983}
Snyder A and Love J 1983 {\em Optical Waveguide Theory\/} (Chapman \& Hall)

\bibitem{Chen-Nielsen-Gregersen-Lodahl-Mork:2010}
Chen Y, Nielsen T~R, Gregersen N, Lodahl P and Mork J 2010 {\em Physical Review
  B\/} {\bf 81} 125431

\bibitem{Russel-Yeung-Hu:2012}
Russell K, Yeung K and Hu E 2012 {\em Physical Review B\/} {\bf 85} 245245

\bibitem{Oulton-Zhang:2009}
Oulton R, Sorger V, Zentgraf T, Ma R, Gladden C, Dai L, Bartal G and Zhang X
  2009 {\em Nature\/} {\bf 461} 629--632

\bibitem{Paulus2001}
Paulus M, Gay-Balmaz P and Martin O 2001 {\em Physical Review E\/} {\bf 63}
  66615

\bibitem{Kottmann-Martin-Smith-Schultz:2001}
Kottmann J~P, Martin O~J~F, Smith D~S and Schultz S 2001 {\em Journal of
  Microscopy\/} {\bf 202} 60--65

\bibitem{ColasdesFrancs-Hugonin:2011}
{Colas des Francs} G, Hugonin J~P and \v{C}tyrok{\'y} J 2011 {\em
  Optical and Quantum Electronics (Proc. OWTNM10)\/} {\bf 42} 557--570

\bibitem{Nauert-Nordlander-Link:2014}
Nauert S, Paul A, Zhen Y~R, Solis D, Vigderman L, Chang W~S, Zubarev E~R,
  Nordlander P and Link S 2014 {\em ACS Nano\/} {\bf 8} 572--580

\bibitem{Jackson:1998}
Jackson J 1998 {\em Classical electrodynamics\/} 3rd ed (Hoboken: John Wiley \&
  Sons)

\bibitem{GCFIJMS:2009}
{Colas des Francs} G 2009 {\em International Journal of Molecular
  Science\/} {\bf 10} 3931--3936

\bibitem{LeRu-Auguie:2013}
Ru E~L, Somerville W and Augui\'e B 2013 {\em Physical Review\/} {\bf 87}
  012504

\bibitem{Grigoriev-Stout:2015}
Grigoriev V, Bonod N, Wenger J and Stout B 2015 {\em ACS Photonics\/} {\bf 2}
  263

\bibitem{Carminati-Greffet-Henkel-Vigoureux:2006}
Carminati R, Greffet J, Henkel C and Vigoureux J 2006 {\em Optics
  Communications\/} {\bf 261} 368--375

\bibitem{JCPGCF:2005}
{Colas des Francs} G, Girard C, Juan M and Dereux A 2005 {\em Journal
  of Chemical Physics\/} {\bf 123} 174709

\bibitem{OptExpGCF:2008}
{Colas des Francs} G, Bouhelier A, Finot E, Weeber J~C, Dereux A,
  CGirard and Dujardin E 2008 {\em Optics Express\/} {\bf 16} 17654-- 17666

\bibitem{Deeb-Bachelot-Plain-Soppera:2010}
Deeb C, Bachelot R, Plain J, Baudrion A~L, Jradi S, Bouhelier A, Soppera O,
  Kain J, Huang L, Ecoffet C, Balian L and Royer P 2010 {\em ACS Nano\/} {\bf
  4} 4579--4586

\bibitem{Zhou-Plain:2014}
Zhou X, Deeb C, Vincent R, Lerond T, Adam P~M, Plain J, Wiederrecht G~P, ,
  Charra F, Fiorini C,  {Colas des Francs} G, Soppera O and Bachelot
  R 2014 {\em Applied Physics Letters\/} {\bf 104} 023114

\bibitem{Khurgin-Sun:2009}
Khurgin J~B and Sun G 2009 {\em J. Opt. Soc. Am. B\/} {\bf 26} B83--B95

\bibitem{Kim-Leung-George:1988}
Kim Y~S, Leung P~T and George T~F 1988 {\em Surface Science\/} {\bf 195} 1--14

\bibitem{GCF-Derom-Vincent-Bouhelier-Dereux:2012}
{Colas des Francs} G, Derom S, Vincent R, Bouhelier A and Dereux A
  2012 {\em International Journal of Optics\/} {\bf 2012} 175162

\bibitem{Koenderink:2010}
Koenderink A~F 2010 {\em Optics Letters\/} {\bf 35} 4208--4210

\bibitem{Derom-Vincent-Bouhelier-GCF:2012}
Derom S, Vincent R, Bouhelier A and {Colas des Francs} G 2012 {\em
  Europhysics Letters\/} {\bf 98} 47008

\bibitem{Sauvan-Lalanne:2013}
Sauvan C, Hugonin J~P, Maksymov I~S and Lalanne P 2013 {\em Physical Review
  Letters\/} {\bf 110} 237401

\bibitem{ZambranaPuyalto-Bonod:2015}
Zambrana-Puyalto X and Bonod N 2015 {\em Physical Review B\/} {\bf 91} 195422

\bibitem{Kristensen-Ge-Hughes:2015}
Kristensen P, Ge R~C and Hughes S 2015 {\em arxiv\/} {\bf arxiv:1501.05938v1}

\bibitem{Jiang-Perrin-Lalanne:2015}
M J~Y, Perrin M and Lalanne P 2015 {\em Physical Review X\/} {\bf 5} 021008

\bibitem{Leung:1990}
Leung P~T 1990 {\em Phys. Rev. B\/} {\bf 42} 7622

\bibitem{Castanie-Boffety-Carminati:2010}
Castanie E, Boffety M and Carminati R 2010 {\em Optics Letters\/} {\bf 35}
  291--293

\bibitem{Girard-Cuche:2015}
Girard C, Cuche A, Dujardin E, Arbouet A and Mlayah A 2015 {\em Optics
  Letter\/} {\bf 40} 2116--2119

\bibitem{anger06a}
Anger P, Bharadwaj P and Novotny L 2006  {\bf 96} 113002

\bibitem{Kim:2015}
Kim J, Song J~H, Jeong K~Y, Ee H~S and Seo M~K 2015 {\em Opt. Express\/} {\bf
  9} 11080--11091

\end{thebibliography}
\providecommand{\newblock}{}

\end{document}